\documentclass[sigconf]{acmart}

\usepackage[english]{babel}
\usepackage{blindtext}

\renewcommand\footnotetextcopyrightpermission[1]{} % removes footnote with conference info
\setcopyright{none}

\usepackage{tikz}
\usepackage{amsmath}
\usepackage[font=small,labelfont=bf,textfont=md]{caption}
\usepackage{subcaption}
\usepackage{multirow}
\usepackage{booktabs}
\usepackage[skip=5pt]{caption}  % Adjust the skip value as needed
\usepackage{xcolor}
\usepackage{verbatim}
\usepackage{enumitem}
\usepackage{hyperref}
\usepackage{breakurl}
\usepackage{xurl}
\usepackage{adjustbox}

\usepackage[table,xcdraw]{xcolor}

\usepackage{listings}
\hypersetup{
    breaklinks=true
}

\usepackage{graphicx}
\usepackage[normalem]{ulem}
\useunder{\uline}{\ul}{}

\usepackage{filecontents}

\usepackage{titlesec}
\titlespacing*{\section}{0pt}{10pt}{5pt}
\titlespacing*{\subsection}{0pt}{5pt}{5pt}
\acmDOI{}

\acmISBN{}

\acmConference{}
\acmPrice{}

\begin{document}
%-------------------------------------------------------------------------------

%don't want date printed
\newcommand{\sys}{\textsf{LLMBridge}}

\newcommand{\fahadprev}[1]{\textcolor{blue}{}}
\newcommand{\fahad}[1]{\textcolor{purple}{}}
\newcommand{\noah}[1]{\textcolor{purple}{#1}}
\newcommand{\abd}[1]{\textcolor{purple}{}}
\newcommand{\hiba}[1]{\textcolor{purple}{}}
\newcommand{\newtext}[1]{\textcolor{red}{#1}}

\newcommand*\circled[1]{\tikz[baseline=(char.base)]{\node[shape=circle,fill=black,text=white,draw,inner sep=1pt,scale=0.75] (char) {#1};}}

\newcommand{\ttimes}{$\times$}

\newcommand{\stype}{\texttt{service\_type}}

\newcommand{\stypes}{\texttt{service\_types}}

\newcommand{\tight}{
  \setlength{\parskip}{0pt}  % Remove space between paragraphs
  \setlength{\topsep}{2pt}   % Adjust space before/after the verbatim
  \setlength{\partopsep}{0pt} % Remove extra space before verbatim
}

\renewcommand{\paragraph}[1]{\medskip{}\noindent{}\textbf{#1}}

\title{\Large \bf \sys{}: Reducing Costs to Access LLMs in a Prompt-Centric Internet}

% \author{Paper \#83, 12 pages body, 15 pages total}

% %for single author (just remove % characters)
\author{Noah Martin$^*$}
\affiliation{Tufts University\country{}}
\author{Abdullah Bin Faisal$^*$}
\affiliation{Tufts University\country{}}
\author{Hiba Eltigani}
\affiliation{Tufts University\country{}}
\author{Rukhshan Haroon}
\affiliation{Tufts University\country{}}
\author{Swaminathan Lamelas}
\affiliation{Tufts University\country{}}
\author{Fahad Dogar}
\affiliation{Tufts University\country{}}

% \renewcommand{\shortauthors}{First Author, et al.}

%-------------------------------------------------------------------------------
\begin{abstract}
%-------------------------------------------------------------------------------

Today's Internet infrastructure is centered around content retrieval over HTTP, with middleboxes (e.g., HTTP proxies) playing a crucial role in performance, security, and cost-effectiveness. 
We envision a future where Internet communication will be dominated by  ``prompts'' sent to generative AI models.
For this, we will need proxies that provide similar functions to HTTP proxies (e.g., caching, routing, compression) while dealing with unique challenges and opportunities of prompt-based communication. 

As a first step toward supporting prompt-based communication, we present \sys{}, an LLM proxy designed for cost-conscious users, such as those in developing regions and education (e.g., students, instructors). \sys{} supports three key optimizations: model selection (routing prompts to the most suitable model), context management (intelligently reducing the amount of context), and semantic caching (serving prompts using local models and vector databases). These optimizations introduce trade-offs between cost and quality, which applications navigate through a high-level, bidirectional interface.

As case studies, we deploy \sys{} in two cost-sensitive settings: a WhatsApp-based Q\&A service and a university classroom environment. The WhatsApp service has been live for over twelve months, serving 100+ users and handling more than 14.7K requests. In parallel, we exposed \sys{} to students across three computer science courses over a semester, where it supported diverse LLM-powered applications --- such as reasoning agents and chatbots --- and handled an average of 500 requests per day.

We report on deployment experiences across both settings and use the collected workloads to benchmark the effectiveness of various cost-optimization strategies, analyzing their trade-offs in cost, latency, and response quality.
% As a first step, we present \sys{}, an LLM proxy supporting the needs of prompt-based communication for cost-conscious users, such as in developing regions. It supports three optimizations: model selection, routing prompts to the most suitable model; context management, intelligently reducing the amount of context; and semantic caching, serving prompts using local models and vector databases.
% These optimizations present tradeoffs in terms of cost and quality, which applications navigate through our high level, bidirectional interface.
% As a case study, we implement a WhatsApp-based Q\&A service that uses \sys{} to provide a rich set of features.
% This service has been deployed on a small scale (100+ users) for 8+ months, and has handled over 14.7K requests so far.
% We report on the experiences of running this service and microbenchmark various cost-optimizations.

\end{abstract}

\maketitle

\def\thefootnote{*}\footnotetext{These authors contributed equally to this work}\def\thefootnote{\arabic{footnote}}

\section{Introduction}

We are transitioning into a \emph{prompt-centric} world:
instead of accessing HTTP content, such as web pages (e.g., \url{www.sigcomm.org}), users are interacting with generative AI tools, especially Large Language Models (LLMs), by crafting prompts (e.g., ``Tell me about Sigcomm'') to retrieve and access information~\cite{perplexityNYT}.
This paradigm shift is reshaping how we consume and produce digital content including web search, ecommerce, translation, and more~\cite{rufusTechCrunch, llmTranslation}, and will have impact on various aspects of the Internet ecosystem, including the role of middleboxes.  

Just as HTTP middleboxes have become a crucial part of today's Internet infrastructure by providing important functions, such as optimizing performance (e.g., compression), ensuring security (e.g., firewalls), and managing traffic (e.g., load balancing), we posit that similar prompt-centric proxies will become mainstream in the emerging Internet infrastructure. 
For instance, a prompt-centric proxy while routing a query to a suitable model may consider network latency, GPU availability, and the quality of the model - much like how CDNs can take into account latency and load. However, the unique opportunities and challenges in prompt-centric communication -- from the flexibility in answering a prompt to changing the prompt itself and the context used for it -- lead to a greater need to make these prompt-centric proxies a first-class citizen of end-to-end communication. We argue that compared to today's Internet, in a prompt-centric world, we would need proxies that can support a wider range of optimizations, requiring a richer set of semantics involving the applications and the proxies.

While the design space for supporting prompt-centric proxies is large, we focus on a proxy that supports cost-saving optimizations.
Unlike typical content access in today’s Internet, accessing LLMs can be very costly~\cite{local_llm}.
For example, OpenAI charges its users \$200/month \cite{openai_pricing} to provide access to their flagship models!
Similarly, hosting a proprietary model on a cloud provider can cost thousands of dollars per month~\cite{llm_economics, llempower}.
As a result, using LLMs could cost as much as 10\ttimes{} a standard Google search~\cite{alphabet_reuters}.  

In this paper, we identify three opportunities for cost optimizations that can be performed by a proxy: model selection, context management, and caching. Selecting the best model (or \emph{model selection}) to handle a prompt can greatly improve the cost, as prices across models can vary by more than 300\ttimes{}.
Judicious \emph{context management} enables smaller prompts, by selecting only the required context needed to answer the query, further reducing cost.
Effective \emph{caching} can potentially bypass (or reduce) LLM calls altogether, thereby reducing cost.

These optimizations have trade-offs that applications need to navigate.
A suitable proxy should allow applications to \emph{delegate} the responsibility of making these decisions via a simple interface that captures the high-level requirements (e.g., cost); in return, the proxy should be \emph{transparent} with the choices it makes (e.g., which model or context was used) to respond to a prompt.
Finally, users should have the option to \emph{iterate} (e.g., ask for a better quality response) if they are not satisfied with the generated response.

\fahad{We should consistently use some term for low-cost/small models here as well as the rest of the paper}
These principles underlie the design of \sys{}, which supports cost-savings optimizations, along with suitable control over these optimizations through a high level, bidirectional API. 
Our design allows simple, well known LLM cost saving optimizations as well as delegation-based ``smart'' strategies that use low-cost models for improved decision making. Specifically, the \emph{model adapter} can use a model to decide which of the LLMs to use, leveraging the strengths of each to consistently deliver responses that are inexpensive and high quality.
The \emph{context manager} leverages a low-cost model to decide what input is necessary for an expensive model.
Finally, the \emph{cache} provides an interface over a vector database to retrieve high quality information.
In addition to supporting low-level semantic similarity based operations, it internally uses a local model to populate the cache and respond to prompts using suitable cached content.

These three building blocks are abstracted by our API, which is centered around high level objectives and bidirectional regeneration of responses.
Applications can \emph{delegate} by specifying a ``service type'' when invoking the proxy, which determines the specific optimizations that should be enabled.
The bidrectional interface provides \emph{transparency} to applications, by returning the details of what configuration is picked (model, context, cache hit/miss) to the caller.
If necessary, callers can then \emph{iteratively} regenerate the response modifying the service type to express updated preferences for these cost saving optimizations.
This allows control over the various inherent trade-offs that are present in these scenarios, while hiding the implementation details of supporting the various optimizations.

As a case study, we build a WhatsApp-based Q\&A service that uses \sys{}.
WhatsApp~\cite{whatsapp} is highly popular in developing regions so providing LLM services over a familiar interface can be useful.
However, the interface also creates new challenges (e.g., message oriented nature) which require additional support from the proxy such as aggressive use of prefetching as well as nudging users to explore cached content through easy-to-navigate buttons. 
We have implemented and deployed \sys{} and our WhatsApp Service on the AWS cloud~\cite{aws} in a serverless environment with a key value store to maintain state.

Our small scale deployment of the Q\&A service has been in production for 12+ months with 100+ users who have sent over 14.7K requests.
Our deployment experience reveals that user prompts range from topics on health and well-being to cultural themes, and are a mix of factual and subjective questions.
Additionally, over this period, the increase in quality of newer LLMs has resulted in cheaper and smaller variants (e.g., GPT4o-mini) being as capable for handling user queries as their expensive counterparts (e.g., GPT4o).
We share these insights, and more, to shed light on optimizations a system like \sys{} can enable based on the interplay between newer LLMs and representative usage scenarios.

To complement this production-facing deployment, we also integrated \sys{} into three university courses, where approximately 60 students used it to build diverse LLM-powered applications.
This classroom deployment provided an opportunity to evaluate \sys{} under instructional constraints --- including limited budgets, varied levels of technical expertise, and sustained development over time.
Over the course of a semester, \sys{} handled roughly 500 requests per day, supporting use cases ranging from multi-agent reasoning to social good chatbots.
By leveraging cost-saving features, total inference costs were kept under \$10 across all three courses.
Post-course surveys further showed that 75\% of students found the system easy to get started with, and over half reported that it was intuitive to integrate into their project workflows.

Together, these deployments offer insights into how \sys{} simplifies the use of LLMs across both real-world and educational settings — enabling cost-efficient operations while remaining easy to use and flexible for a wide range of users.

\fahad{Will be good to have the actual insights rather than just saying we did XYZ.}
In addition to these case studies, which show the feasibility of supporting realistic workloads over \sys{}, we also evaluate specific cost optimization strategies presented in this paper.
Our results in \S\ref{sec:evaluation} demonstrate cost reduction strategies with savings of over 30\%.
Specifically, we evaluate combining multiple models to answer queries in order to leverage cheaper ones as much as possible.
We test an intelligent context management strategy that reduces input tokens based on the conversation history we gather from the WhatsApp service.
Lastly, we demonstrate the effectiveness of caching information relevant to the WhatsApp queries in order to reduce calls to expensive LLMs. 
%While the cost savings are significant, they also present %time/quality trade-offs. 
%The API for \sys{} is designed to navigate these tradeoffs.

Overall, we make the following contributions in this paper:
% \fahad{Maybe pitch the "Case for a proxy -- section 2" as a contribution??}
%\paragraph{Contributions:}

\begin{comment}

\begin{itemize}[leftmargin=*]
\item Make a case for an LLM proxy that supports several cost-optimizations for model selection, context management, and caching, and provides control over them through a suitable interface. 
\item Design of \sys{}, showing how its three components enable existing and smart optimizations, along with a high level, bidirectional API that provides control over the various optimizations, alongside a lightweight serverless deployment on AWS.
\item Share deployment experiences from both a WhatsApp Q\&A service (100+ users, 8+ months) and classroom settings (60 students, 75K requests), highlighting real-world and instructional usage.
\item Evaluate strategies for cost reduction on real user workloads to demonstrate their cost/quality trade-offs.
\end{itemize}
\end{comment}

\begin{itemize}[leftmargin=*]
\item Make a case for an LLM proxy that supports cost-saving optimizations — including model selection, context management, and caching — and provides control over them through a suitable interface.
\item Present the design of \sys{}, showing how its three components enable both standard and smart optimizations via a high-level, bidirectional API, and implement it as a lightweight, serverless deployment on AWS.
\item Share deployment experiences from a WhatsApp-based Q\&A service (100+ users, 12+ months) and university classroom settings (60 students, 75K requests), both deployed in cost-sensitive environments, highlighting real-world and instructional usage.
\item Evaluate cost-optimization strategies on real workloads, demonstrating trade-offs in cost, quality, and latency.
\end{itemize}

The rest of the paper is organized as follows: In~\S\ref{sec:motivation}, we discuss the case for a proxy supporting various cost optimizations and how it can support a wide range of LLM applications, drawing analogies to HTTP proxies and highlighting unique challenges.
In~\S\ref{sec:design}, we present the design of \sys{} including its API and details on the model selection, context management, and caching process.
We describe our implementation of the proxy in~\S\ref{sec:implementation}.
We share our deployment experiences and microbenchmarks in~\S\ref{sec:evaluation}.
Finally, we note that given the importance of LLMs, there is an increasing body of relevant work, both on abstractions/middleware (e.g., LangChain~\cite{langchain}, Bedrock~\cite{bedrock}) as well as specific optimizations, such as model routing (e.g., HybridLLM~\cite{ding2024hybrid}, RouteLLM~\cite{ong2024routellmlearningroutellms}) and semantic caching (e.g., GPTCache~\cite{bang-2023-gptcache}).
These concurrent proposals compliment and reinforce the need to consolidate these important optimizations in a proxy, potentially benefiting from \sys{}'s design.
We elaborate on these works in~\S\ref{sec:related-work}.
% various aspects of the \sys{} design: the optimizations can fit into the overall proxy design of \sys{}.
% The focus of other abstractions is on different aspects of LLM usage whereas \sys{} focuses on cost-optimizations. We elaborate on these works in~\S\ref{sec:related-work}. 

%In \S\ref{sec:related-work}, we review related work including %concurrent work such as model routing %(HybridLLM~\cite{ding2024hybrid}, %RouteLLM~\cite{ong2024routellmlearningroutellms}) and semantic %caching (GPTCache~\cite{bang-2023-gptcache}). These optimizations %are complementary to the ones we implement in \sys{} and also fit %into the overall proxy design.
% \fahad{We need an additional line or two capturing the related work and placing us in a suitable way. Also, use higher level terms like model routing and semantic cache, which are more popular, and give these projects as specific examples}

\paragraph{Ethical Concerns:} All data collection and analysis is carried out in compliance with our university Institutional Review Board (IRB) process and is covered by the terms and conditions and privacy policy accepted by the users.

\section{Motivation}\label{sec:motivation}

% \noah{Have a Table of middlebox use cases, and what challenges there would be with LLMs. PEPs (compression) particularly important for developing regions. Caching, Routing. Load Balancing (CDN). Content filtering - block based on URLs with a denylist. How does that work with LLMs? Can refer to related work on LLM content filtering. They might not be in open source models - need it in the proxy. Principles that lead to the current design: ie variable quality. Input can also be modified. 3 key differences: Output quality can vary. Input can be modified. Response can be gathered iteratively.}

In this section, we first make the case for LLM proxies playing an important role in the emerging prompt-centric Internet.
We then focus on a specific use for these proxies: to support cost-optimizations related to model selection, context management and caching.

\begin{table*}[]
\footnotesize
\begin{tabular}{|l|l|l|}
\hline
\textbf{Use Case}                                                    & \textbf{HTTP}                                                                                                               & \textbf{LLM}                                                                                                                                                                                                                                              \\ \hline
\begin{tabular}[c]{@{}l@{}}Content\\ filtering\end{tabular} & Domain denylist                                                                                                    & \begin{tabular}[c]{@{}l@{}}Filter based on the input prompt or contents of the output \\ Limit model access to specific users\end{tabular}                                                                                                             \\ \hline
Compression                                                 & \begin{tabular}[c]{@{}l@{}}Precompute Javascript (e.g., Prophecy~\cite{prophecy})\\ Webpage compression (e.g., Flywheel~\cite{flywheel})\end{tabular}          & \begin{tabular}[c]{@{}l@{}}Context can be minimized through truncation, summarization, or RAG\\ Reducing number of examples provided by few shot prompting\\ These reduce the request size but also may reduce quality of responses\end{tabular} \\ \hline
Routing                                                     & \begin{tabular}[c]{@{}l@{}}CDNs select server based on content\\ availability, load, proximity, etc\end{tabular} & \begin{tabular}[c]{@{}l@{}}Selecting from various models can reduce response time, mitigate\\ datacenter load, and affect cost. However, smaller models can\\ also reduce response quality. Model routing can combine \\ results of multiple models in series or parallel.\end{tabular}                                         \\ \hline
Caching                                                    & \begin{tabular}[c]{@{}l@{}}URL matching with aliases (e.g., RC2~\cite{remote-control-caching})\\ Prefetching predicted content (e.g., Marauder~\cite{marauder})\end{tabular} & \begin{tabular}[c]{@{}l@{}}Augment cached data with local LLM to make it relevant for new queries.\\ Prefetch prompts related to most recent queries.\end{tabular}                                                                             \\ \hline

\end{tabular}
\caption{Different use cases, and how they apply to HTTP proxies vs. LLM Proxies}
\label{proxy-use-cases}
\end{table*}

\subsection{A Case for LLM Proxies}
HTTP proxies provide numerous features in today's Internet, from security and access control~\cite{durumeric2017security} to performance enhancements, such as compression, load balancing and caching \cite{marauder, remote-control-caching, yodaLoadBalancer}.
These operate on assumptions of the content retrieval nature of modern Internet traffic, such as caching the data represented by a URL.
Proxies are particularly important for developing regions where bandwidth and energy is costly and low-end devices are the norm~\cite{flywheel, flexiweb, shandian, prophecy}.
As more and more web features adopt LLMs, we will find ourselves needing similar features (if not more) from proxies that can operate on the new prompt-centric network traffic. While proxies have their own drawbacks (e.g. new failure modes and lack of mobility), there have been efforts in the networking community to address these~\cite{tapa}.

In Table~\ref{proxy-use-cases}, we summarize a few common HTTP proxy use cases that have analogies to LLMs and what new challenges (and opportunities) the LLM version would face. For example, instead of denylisting URLs to filter content, LLM proxies would need to handle ways users can craft prompts to bypass LLM restrictions. Even if one provider implements content filtering, a proxy may still be necessary to filter content from a collection of model providers and for specific uses (e.g., schools).
Bandwidth and latency reduction through compression and prefetching is also common in proxies~\cite{flywheel, prophecy}.
This too can be extended to LLMs, by reducing the data needed for prompts.
However, there is an additional challenge with applying this to LLMs because modifying the context, including summarization or limiting the number of examples, has an effect on output quality.
Lastly, while typical HTTP traffic benefits from proxies that provide routing functions, LLMs can benefit from ``model routing''.
An LLM proxy can route prompts not only based on proximity and load but also quality and cost of available models or even route queries to multiple models. 

\fahad{strengthen this paragraph by adding a few more lines describing how putting it closer to the user vs the model could provide different kinds of benefits.}
There are many advantages to placing these features in a proxy at a nearby cloud (edge) location rather than a local application library. For instance, low-powered IoT devices benefit from avoiding local context storage and from not running optimizations involving additional models. Access control and rate limiting for certain models can also be implemented when provider API keys are not needed on clients, we present a use case for this in~\S\ref{sec:course-projects}.
A proxy that is aware of which cloud regions have available LLMs can also account for the sparsity of datacenters in developing regions~\cite{divide-conext} and the limited model support in some locations~\cite{bedrock_availability, azure_availability}.
All while benefiting from stable network conditions between clouds when deciding the best LLM for a task~\cite{cloud_paths_www}.

Many features that are desirable to have in LLM proxies exhibit trade-offs such as increased quality at the expense of higher latency. We identify three properties that highlight the differences between an LLM proxy and traditional middleboxes:
\begin{enumerate}
\item Output quality can \emph{vary} depending on the exact inputs and also which LLM is used (e.g., more or less parameters in the model).
\item Unlike most HTTP requests, the input to LLMs are \emph{flexible} and can be modified to produce similar results.
\item Responses can be generated \emph{iteratively} to refine and improve the result, particularly useful for emerging LLM agents. 
\end{enumerate}

% \fahad{We should highlight at least two things: 1) LLM Proxies are going to be making more important functions than traditional proxies so the need to make them visible and involve them in supporting applications semantics is even more critical (e.g., are they allowed to change the prompt or use any model they have, etc) (Basically we need to touch upon hidden vs visible proxies), ii) Many of the decisions they will make will often involve trade-offs, and application/user is typically in the best position to decide on that} 

These principles motivate the need to consider these proxies as a first-class citizen in the end-to-end communication: they need to be visible to the applications and must work together with the applications to navigate the various trade-offs. This is similar in spirit to the various cases to make hidden middleboxes in today's Internet visible to applications~\cite{doa, tapa, httpnarrowwaist}. Prompt-centric proxies may have application specific needs for control and transparency as well, such as strict data governance and auditing that are likely to apply to healthcare applications using LLMs. We believe the need for proxies and applications to work together is even greater for prompt-centric communication and the right time to consider these issue is \emph{now} rather than when these proxies are already widely deployed and used in ad-hoc ways. 

%Unlike many HTTP middleboxes that can be deployed without application knowledge, in our %case the application must work with the proxy to reach the right trade offs concerning %cost/quality/time. This requires the use of an expressive interface, discussed in detail %in~\ref{sec:api_design}. 

% \fahad{This paragraph has some useful points but I don't think they belong here. Will be best to discuss this earlier when we talk about proxies and optimizations. Perhaps when we are discussing the table.}
% Navigating these trade offs could be implemented locally, as an application library, but there are many advantages to placing them in a proxy, which is hosted in a nearby cloud (edge) location. For instance, low powered IoT devices benefit from not storing context locally nor running optimizations involving additional models. A proxy that is aware of which cloud regions have available LLMs can benefit from stable network conditions between clouds when deciding the best LLM for a task~\cite{cloud_paths_www}. Of particular interest to us is how the tradeoffs can minimize cost.

% \fahad{May want to update/verify current costs. Also, may want to say that the cost of a models are going down, usually when newer models are released but the overall cost is still high, especially for using the state-of-the-art models}

\subsection{Cost Saving LLM Proxy}
\label{sec:cost-saving-techniques}
%Cost could be an important consideration for accessing LLMs. 

%Table~\ref{cost-table} lists example use case costs. While small individually, in the %aggregate these costs can add up to a prohibitively large bill for some users.

While the design space for LLM proxies is large, in this paper, we focus specifically on proxy functionally that can reduce the \emph{cost} of accessing LLMs. These optimizations have broad applicability, but are particularly useful for cost-conscious users, such as those in developing regions, researchers, students, etc -- anyone who cannot afford to always use the highest quality (and highest cost) model available in the market. As prior work in developing regions has shown, users in these areas are particularly cost sensitive and are willing to make various compromises if that results in lower cost~\cite{missit}. 

The cost of an LLM typically varies across models and depends on the number of input and output \emph{tokens}, with one word being roughly 1.3 tokens~\cite{openai_docs}. Generally, output tokens cost more than input tokens (5\ttimes{} difference for Claude 3 models~\cite{anthropic}). As new models are released that are cheaper than their predecessors, state of the art models remain costly.
% For example, Claude Opus output tokens are 375\ttimes{} as costly as Amazon Titan Text Lite~\cite{bedrock_pricing}. 
For example, GPT-4.5 is 250\ttimes{} as costly as GPT4o-mini~\cite{azure_openai_pricing}. 

In this paper, we focus on how well known optimization techniques can fit into the design of a proxy, specifically the API, as well as evaluate common techniques on real world datasets from our WhatsApp deployment~\ref{sec:deployment} which gives insights into their applicability~\ref{sec:evaluation}. Next, we review three areas of cost-optimization.
% can provide cost reducing features in a proxy.

\begin{comment}
\begin{table}
\footnotesize
\begin{tabular}{|l|l|l|l|}
\hline
Use case                & Titan Text Lite & Haiku     & Opus    \\ \hline
1M output tokens           & \$0.2           & \$1.25    & \$75    \\ \hline
Writing 5000 token lecture & \$0.00075       & \$0.00125 & \$0.375 \\ \hline
Using full context window  & \$0.0006        & \$0.05    & \$3     \\ \hline
\end{tabular}
\caption{Cost of LLM use cases (in USD)}
\label{cost-table}
\end{table}
\end{comment}

% \fahad{The following subsections can be shrinked a little. May even be demoted to bold headings rather than subsections. May briefly mention/acknowledge latency/other benefits too, on top of cost}

\paragraph{Model selection.}
\label{sec:motivationmodelselection}
The most important aspect in determining the cost of an LLM task is typically the choice of model, which affects a number of factors like quality of responses, cost and latency.
Some APIs integrate with multiple providers~\cite{langchain}, but it is not always clear \emph{how} to select a model.
Other frameworks are restricted to only some models - for example, OpenAI’s Assistants API~\cite{openai_assistant} does not work with smaller models (e.g., Phi~\cite{phi3}).
% \footnote{Model latency is determined by time to first token (TTFT) and tokens per second (TPS). The total time to generate a response is $\text{TTFT} + number\_of\_tokens / \text{TPS}$.
% Models with more parameters typically take more time to generate tokens (lower TPS), and cost more.}.
% Some frameworks that manage conversations with LLMs are restricted to only the models supported by the API provider - for example, OpenAI’s Assistants API~\cite{openai_assistant} does not work with smaller models (e.g,. Phi~\cite{phi3} ).

The key observation underlying this optimization is that most expensive models can be an overkill for certain, easier, queries.
Therefore, an intelligent strategy for picking an appropriate model, also referred to in recent work as model routing~\cite{frugalgpt, ong2024routellmlearningroutellms, ding2024hybrid}, may significantly reduce costs while maintaining the quality of the most expensive model.

\paragraph{Context Management.}\label{sec:motivation_context}
The amount of context provided to a model affects the number of input tokens, and therefore cost, as well as the latency to process the input tokens.
One strategy for chat applications, ``last-k'', provides the $k$ previous messages as context for the next message.
% Another technique for adding context to improve quality is retrieval augmented generation (RAG)~\cite{lewis2021retrievalaugmented}. Text that may be relevant to the query is split into documents of a fixed length, and an embedding is created for each document. At generation time, the LLM input is augmented with the documents that have a similar embedding using a metric such as cosine distance or retrieval models~\cite{khattab2020colbert}.
A larger $k$ may increase response quality by including more relevant information, at the expense of higher cost. Additional strategies exist to improve quality without providing all possible context~\cite{ragcache, RAG, TCRA}, and we explore how \sys{} can support these in~\S\ref{sec:design_context_manager}.
 % Returning to the example of the Assistants API~\cite{openai_assistant}, developers can add context to requests in the form of extra ``documents'' with control over the maximum number. However, this API does not make it clear that this can directly affect cost nor how to pick a suitable value.

% \fahad{I think this para can be removed}
% Attaching more context (i.e., input tokens) to a request can also increase latency. While this latency increase is small compared to increasing the output tokens, it can have a significant effect on the TTFT of some queries~\cite{openai_latency}. This suggests context management strategies that reduce cost without sacrificing on quality could also reduce latency.

\begin{figure}[!t]
    \centering
    \subfloat[Input tokens of context strategies]{\includegraphics[width=0.48\columnwidth]{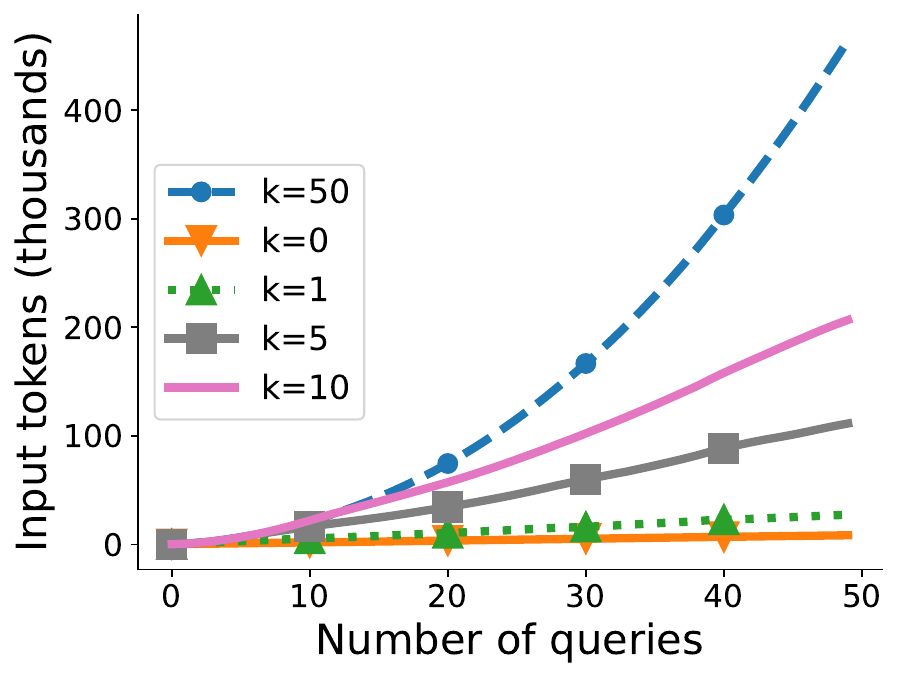}
    \label{subfig:cost_motivation}}
    \subfloat[Quality of context strategies]{\includegraphics[width=0.48\columnwidth]{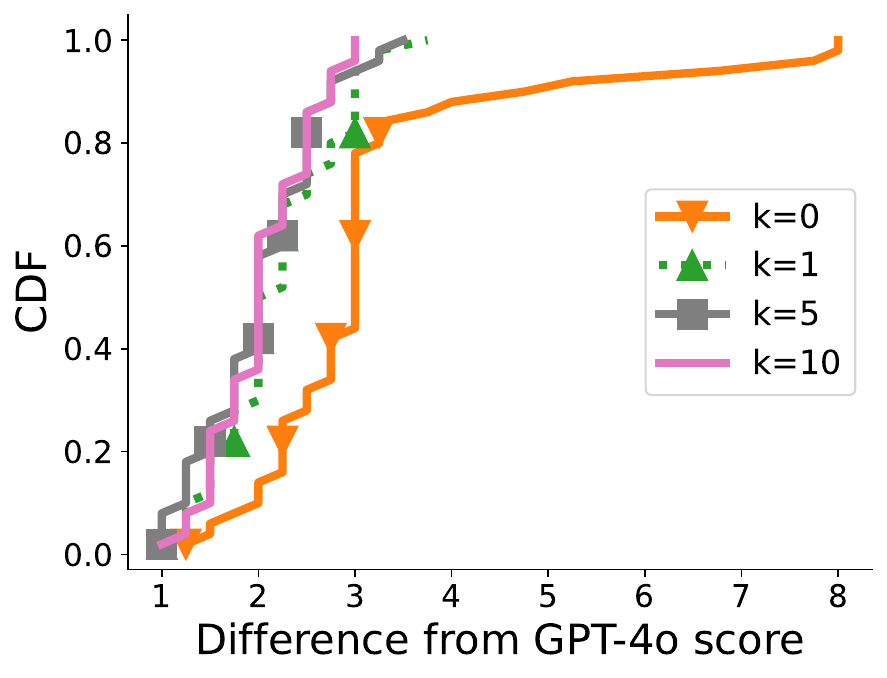}
    \label{subfig:quality_motivation}}

    \caption{\ref{subfig:cost_motivation} Compares the cost, measured by input tokens, when various amounts of previous messages ($k$) are in the the context. \ref{subfig:quality_motivation} Compares the quality of each strategy with $k=50$ as the reference.}
    \label{fig:motivation_context}
\end{figure}

To motivate potential cost saving strategies of a proxy, we evaluate cost and response quality using 5 values of $k$ for ``last-k'' in a 50 query conversation from our WhatsApp deployment (\S\ref{sec:deployment}). If we assume all $N$ queries have the same number of input and output tokens, $I$ and $O$, then the input tokens used with $k=N$ is: $I * N + (I + O)N(N-1)/2$.
This is $O(n^{2})$, and Fig~\ref{subfig:cost_motivation} shows that including all context (k=50) grows quadratically while $k=0$ grows linearly.
The maximum context conversation uses 55\ttimes{} the input tokens of no context and $k=1$ is only a 3\ttimes{} increase. The quality of these conversations judged against using full context is shown in Fig~\ref{subfig:quality_motivation}.
Each response is given a score (S) from GPT-4o which is averaged over four runs.
Although using no context is the lowest quality, for this workload of real WhatsApp queries the difference is most evident only in the tail 20\% of messages.
This motivates a simple strategy -- using context only when necessary to substantially improve quality.
We present this strategy in \S\ref{sec:design_context_manager}.

\begin{comment}
Analytically, the number of input tokens used by $N$ queries is:

\begin{equation}
\sum_{i=0}^{N}(I_{i} + \sum_{j=i-k}^{i-1}(I_{j} + O_{j})) \notag
\end{equation}
Where $I_{i}$ and $O_{i}$ are the number of input and output tokens for the $i$th message, respectively, and any negative index message is 0 tokens. For the case of $k=N$ with the simplifying assumption of all messages having the same input and output tokens, $I$ and $O$, the result simplifies to:

% $I \sum_{i=0}^{N}(1 + \sum_{j=0}^{i-1}) + O \sum_{i=0}^{N}\sum_{j=0}^{i-1} 1$

% $I*N + I \sum_{i=0}^{N-1}i + O \sum_{i=0}^{N-1}i$

% $I*N + I * N(N-1)/2 + O * N(N-1)/2$

\begin{equation}
I * N + (I + O)N(N-1)/2 \notag
\end{equation}
%\fahad{Can we just show the final expression under the simplifying %assumptions? If we want the earlier experession it could possible %come as a footnote. We need to convey the quadratic nature of the %expression without making the reader put too much effort in %understanding the analytical expression.}
%
\end{comment}

\paragraph{Caching.}\label{sec:motivation_caching}
This is a well-known strategy for lowering costs and latency by reducing reliance on an expensive or distant resource.
With prompts there is similar potential: a suitable cache can eliminate the need to use an LLM altogether, or cached information can be used to supplement a cheaper LLM to obtain high-quality information at a lower cost. %, the best case for cost and latency.

% In a typical cache, request/response pairs are stored and the system checks if new requests have a cache ``hit'' before performing a more expensive operation.

A traditional HTTP cache is based on exact matches.
This, however, can be limiting for prompts since natural language queries may be \emph{semantically} similar instead (e.g., ``Tell me about SoCC'' vs. ``Talk to me about the SoCC conference'') and can use the same cache entry.
One way to support this is with embeddings of the input text~\cite{ragcache, meancache}.
Embeddings with a high similarity can be considered cache hits.
Since computing embedding requires less resources than generating the response, it may be appropriate to compute locally which enables applications to use the closest cache match when LLMs are unavailable.

An LLM cache can also be populated with high quality data for a cache-local model to construct its responses from.
% This is a form of retrieval augmented generation~\cite{RAG, phi3_rag} where the documents used to augment a query come from cached requests.
Another well known systems technique, pre-fetching, is enabled by this caching approach.
Requesting additional information related to a relevant topic from a high quality model (e.g., follow-up questions, reasoning-chains~\cite{sleep-time-compute}) populates the cache, which can then be used by a local model to generate responses for subsequent queries.

\section{Design}\label{sec:design}

\definecolor{goodgreen}{rgb}{0.3, 0.7, 0.3}

\begin{table}
\begin{tabular}{l}
\cellcolor{gray!20}\texttt{\textbf{request}(\textcolor{goodgreen}{prompt},\textcolor{red}{service\_type},\textcolor{blue}{\{key:value..\}})} \\
\footnotesize \texttt{\# sends request to \sys{}. The result contains the} \\
\footnotesize \texttt{\# response as well as metadata} \\
\\
\cellcolor{gray!20}\texttt{\textbf{regenerate}(\textcolor{red}{service\_type},\textcolor{blue}{\{key:value..\}})} \\
\footnotesize \texttt{\# regenerates the response using either a different} \\
\footnotesize \texttt{\# service\_type or the same one} \\
\hline
\end{tabular}
\caption{\sys{} API}
\label{tab:interface}
\end{table}

\fahad{we need to clearly specify the API somewhere.If it's in Figure 3a, we need to explicitly refer to it in the API section. And perhaps highlight that in the Figure as well}
% \fahad{Figure looks good. We may consider adding applications along with the user including WhatsApp? Also, more models on the right including custom models. Also switch the query (top), response (bottom). }

\begin{figure}
\begin{center}
\includegraphics[width=\columnwidth]{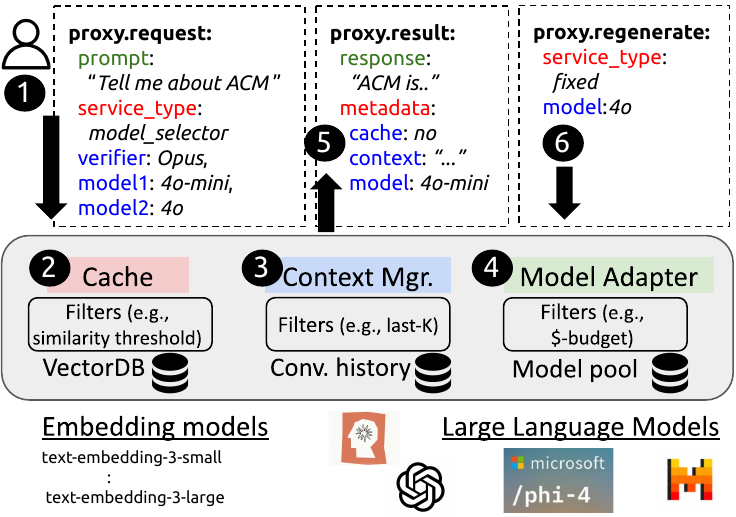}
\caption{Overview of \sys{} design.}
\label{fig:design_overview}
\end{center}
\vspace{-1.5em}
\end{figure}

The properties of prompt-centric communication map to the following design requirements that a cost-saving proxy must satisfy:
\abd{TODO: be consistent with users or applications. Prompts or queries.}
% Generating responses to prompts can be a rich function of the LLMs that are considered, using cached information, the amount of context that is used etc.
% These principles have the following design implications:
% \underline{unified} \underline{high-level} \underline{bi-directional}
% These principles have the following design implication: need for a \underline{unified high-level bi-directional} interface.

% \abd{maybe remove solution from these three points and only discuss requirements.}
\begin{itemize}[leftmargin=*]
    \item \textbf{Simplifying navigating the trade-off space.}
    The \emph{combined} effect of different optimizations on cost and quality are unknown to the applications a priori.
    The proxy should offer a \emph{high-level} interface to capture application preferences.
    High level preferences can be mapped to a \emph{myriad} of low-level optimizations (e.g., model and context selection strategies).
    Barring simple scenarios, it can be challenging to come up with suitable mappings.
    To address this, the proxy should support \emph{delegation} of responsibility.
    % This can be be achieved using intelligent internal small language models.

    \item \textbf{Providing transparency.}
    The proxy should inform applications how their prompts were resolved (e.g., whether a cached response was served), particularly in scenarios involving delegation.
    This is similar to HTTP proxies which insert additional information in the response (e.g., \texttt{X-Cache}) to indicate how they processed the request.
    % Transparency may be crucial for regulated services where decision-making must be auditable (e.g., finance, healthcare).
        
    \item \textbf{Allowing adjustments.}
    Ultimately, applications are best positioned to assess the quality of the generated responses.
    In addition to providing transparency, the proxy should allow applications to refine the responses in an \emph{iterative} fashion.
    
    % When delegating responsibility to the proxy, applications should be informed 
    % , and offer adjustments through an \emph{iterative} process.

\end{itemize}
\subsection{Overview}

% \fahad{First para should introduce the proxy referring to the figure. Its placement -- between applications -- and models. And how it takes common functionality out of the applications, and provides a common interface, but the main focus is on cost optimizations. Second para could introduce the key features of the proxy, similar to the intro. 
% Third and subsequent paragraphs could talk about how applications use the proxy -- it can start with the API. I would suggest taking two running examples -- one that is straightforward and doesn't use the smarts while the other one does, to make the contrast. You already allude to that when you refer to low level vs high level, etc.} 

As shown in Fig~\ref{fig:design_overview}, \sys{} is a proxy that sits between users or applications (e.g., chatbots) and the various LLMs that are available, including proprietary (e.g., OpenAI, Claude), open-source (e.g., Llama, Mistral, DeepSeek), and custom models.
It offers a high-level bi-directional interface (\S\ref{sec:api_design}) that captures application preferences (\circled{1}), while ensuring transparency (\circled{5}) and enabling iterative refinement (\circled{6}) of prompt resolution.

% Under the hood, \sys{} consists of three building blocks (Model Adapter, Context Manager, and Cache) that allow various cost-saving optimizations to be (approximately) implemented~\cite{a,b,c}.
% These building blocks offer low-level control via a ``filters''  based abstraction (e.g., similarity threshold for Cache).
% More importantly they offer \emph{delegation} by using internal task-specific models to decide how to use filter parameters/results based on the given query.

Under the hood, \sys{} consists of three building blocks: the Model Adapter~(\S\ref{sec:design_model_adapter}), Context Manager~(\S\ref{sec:design_context_manager}), and Cache~(\S\ref{sec:design_cache}), which enable various cost-saving optimizations (\S\ref{sec:cost-saving-techniques}) through a filter-based abstraction.
The filters provide low-level control (e.g., setting thresholds) but, more importantly, facilitate \emph{delegation} of responsibility by incorporating low-cost models.

% The filters offer low-level control (e.g., similarity threshold for Cache) but more importantly, to facilitate \emph{delegation} of responsibility, also provide filters based on internal task-specific models as well as a combination of the two.
% The internal task-specific models optionally process the filtered results (e.g., rewriting cached responses).

% To facilitate delegation, the filters are based on internal task-specific models but also offer low-level control  and a combination thereof.

% low-level (e.g., similarity threshold for Cache) control.
% More importantly, they enable \emph{delegation}, where internal task-specific models determine appropriate filter parameters, 

% the cache might be queried after determining no additional context is available. 

% handle the subsequent processing of the results, which may involve additional construction or refinement after the initial filtering step.

% The three primary components of \sys{} are the model adapter, context manager, and cache.
The Model Adapter can \emph{route} prompts to different LLMs and \emph{combine} multiple models, based on their respective abilities (e.g., supported context window) and usage cost.
% to offset some work from more expensive LLMs.
The Context Manager retrieves relevant context (e.g., previous message) to supplement prompts, and supports a number of ways to select context including using a low-cost model to reduce the amount of context sent to expensive LLMs.
Lastly, the cache stores data (e.g., previous prompts, high-quality information) that could help in replying to new prompts - using a low-cost model to turn cached data into suitable replies.

\sys{} uses the \texttt{service\_type} accompanied with each request to decide the order in which to call these components.
Different \stypes{} can map to different orderings.
For all the \stypes{} included in this paper, the order \circled{2}-\circled{4} as shown in Fig~\ref{fig:design_overview} is followed.
We now describe the API and the individual components in more detail.

% For example, with the \texttt{service\_type} set to quality, the context may be retrieved before performing a cache look-up, in order to get a context-aware cached response (higher quality but slower).
% Alternatively, for a cost \texttt{service\_type}, the cache can be queried first to possibly avoid a context lookup, as steps  show.
% The mapping from \texttt{service\_type} to 

% also implementing common features that applications would otherwise need to implement on their own.  
% Such features include handling multiple model formats, which are hidden by the proxy API, and managing conversation context. The focus of our design is chatbot style applications which often maintain a history of the conversation to be used as context. 

%\sys{} is designed to make use of these features to provide cost %optimizations.

% \fahad{This paragraph should reuse some of the stuff from the intro, put in a slightly different way. We should clearly highlight that each module can support a range of optimizations while also mentioning one smart strategy that we use with the help of an LLM}

\subsection{API}\label{sec:api_design}

%\fahad{Do we want to define the API somewhere -- in addition to the figure which I %don't think captures all the details of the API?}
The API for \sys{} is shown in Table~\ref{tab:interface}.
It allows applications to specify a \texttt{service\_type} as part of the request (\texttt{proxy.request}) which maps to a particular configuration of each internal component.
The API is designed to work iteratively, using a bi-directional interface where the proxy responds (\texttt{proxy.result}) with details of the settings that were ultimately used and applications can regenerate responses (\texttt{proxy.regenerate}) using a different \texttt{service\_type}.

\paragraph{How does the API enable delegation?}
The \texttt{service\_type} supports both low-level and high-level specifications, offering a spectrum of delegation of work --- from none to a high degree.
% These are captured by \texttt{service\_types}, offering a spectrum of delegation --- from none to a high degree.

An application can use basic \texttt{service\_types}, specifying low-level parameters (e.g., model to be used).
More importantly, the API exposes more powerful \texttt{service\_types} which delegate these choices to the proxy, which in turn uses low-cost models within each component to decide (e.g., route prompts to LLMs based on complexity).
\fahadprev{An example here would be useful}
Some \texttt{service\_types} require additional parameters, which can be specified as key-value pairs.
We start with basic \texttt{service\_types} and progressively move to higher levels of delegation:

% Different configurations for \sys{}'s components are mapped to different \texttt{service\_type}'s.

%\fahad{In general this section needs more examples, to make things concrete as %well as to show that these features are applicable in a wider range of settings}
% There are basic \texttt{service\_type}'s corresponding to three performance indicators of LLMs.
\begin{itemize}[leftmargin=*]
    \item \texttt{fixed}: Uses a fixed configuration for model, context, and cache.\\
    This can be specified as: \texttt{(model=modelID, cache=skip,..)}

    \item \texttt{quality}: Uses the most expensive model and as much context as the model window allows
    % Only considers the cache if a highly similar cached prompt is present
    % % Optionally, an accuracy metric (e.g., F1 score) can be specified to denote a minimum quality requirement.

    \item \texttt{cost}: Uses the cheapest model with no context

    \item \texttt{model\_selector}: Employs the model selection strategy detailed in \S\ref{sec:design_model_adapter}, initially using a cheaper LLM and falling back to an expensive one if the response quality is below a threshold.
    It uses 5 previous messages (i.e., prompt-response pairs) as context to avoid low quality responses, while keeping the cost down.
    % This is evaluated in \S\ref{sec:evaluation_model_selection}.
    % The optional values allow applications to specify the models to be used in the cascade.

    \item \texttt{smart\_context}: Uses a small model to determine whether the context should include the last five messages or none. This is useful for applications aiming to reduce input token costs while accepting a potential quality trade-off due to false positives, as explained in \S\ref{sec:design_context_manager}.
    % The effectiveness of this approach is evaluated in \S\ref{sec:evaluation_context_manager}.
        
    \item \texttt{smart\_cache}:
    Uses small model to determine whether a prompt can be answered using cached information.
    If there is a cache hit, the small model is used to reply to the prompt given the extra information in the cache. This is described in \S\ref{sec:design_cache}.

    % \item \texttt{speed.} Performs a cache look-up prior to invoking the fastest model with one previous request-response as context.

    % \item \texttt{fixed\_model} uses the LLM specified in the \texttt{hints} parameter.
\end{itemize}

% The highest quality model cannot necessarily be picked a priori since there is not a strict quality ordering, but a suitable default can be chosen and more specific service types used if applications require them.

% Importantly, \sys{} supports \emph{delegation}, leveraging the cost-saving features of each component alongside internal models.

% \begin{itemize}[leftmargin=*]

%     % \item \textit{model\_cascade.} Uses the model cascading strategy detailed in \S\ref{sec:design_model_adapter} to first use a cheaper LLM and only use an expensive LLM if the response quality is poor (below a threshold).

% \end{itemize}

%\fahad{consider something similar to the table we had for Hotnets to discuss use-%cases if we have space?}

%\input{use-case-table}

\paragraph{Transparency.}
While delegating responsibility relieve applications from specifying low-level options (e.g., which LLM to use), they make the prompt-resolution process opaque; applications don't know \emph{how} their prompt was handled.
Similar to HTTP proxies which disclose request resolution details (e.g., \texttt{AGE} and \texttt{X-Cache} for cached content), \sys{}'s responses include \emph{metadata} to provide transparency.
The metadata captures the low-level choices made by each component on behalf of the application, including the model(s) used, the amount of context added, and whether the response was returned from the cache.
This can assist applications wanting to refine responses iteratively, as we discuss next.
% This provides transparency over the choices made by the proxy.
% This enhances transparency into the proxy's decision-making process.
% The metadata consists of different low-level options each component took on behalf of the application, such as the model(s) used, the amount of context added, whether it is a cached response etc.

\paragraph{Iterative nature.}
\sys{} makes iterative refinement a first class concept, as a way to provide finer control to applications.
This is based on the fact that ultimately applications are best suited to asses the quality of the responses.
\sys{} allows applications to regenerate a response, either using the same \stype{} or a different one, based on the information in the metadata provided.
Using the same \stype{} in the regenerate request will nudge the proxy to prioritize quality over cost.
For example, for the \texttt{smart\_context}, regenerating a response entails using more context.
For different \stypes{}, the implementation of \texttt{proxy.regenerate()} can be different.

% update the \stype{} or the provided \texttt{values} to regenerate a response based on the metadata provided.

The iterative nature of prompt resolution is a natural fit for many LLM applications (e.g., ChatGPT) which already include an option in the user interface to regenerate responses or provide other forms of feedback~\cite{llm-in-ui, ai-for-misinformation}.
For example, the WhatsApp Q\&A service~(\S\ref{sec:deployment}) we have built on top of \sys{} contains a ``Get Better Answer'' button for every message.
When pressed, the application signals the proxy to regenerate the response using a higher cost model.

\subsection{Model Adapter}\label{sec:design_model_adapter}

\begin{comment}

\begin{figure}[!t]
\begin{center}
\includegraphics[width=\columnwidth]{figures/verificationstrategysmaller.png}
\caption{Design of our model selection strategy.}
\label{fig:verificationstrategy}
 \vspace{-1.5em}
\end{center}
\end{figure}

\end{comment}

\fahadprev{We should clearly show the the model adapter API somewhere}
The model adapter provides two functions: a unified interface that wraps calls to third party LLMs (which may have different APIs, formats etc.) and a way for applications to delegate the choice of the LLM used.
% The unified interface is intended to hide provider specific details such as formatting of message history, streaming tokens, and response formats (json/text).
% It accepts parameters to specify each of these capabilities, and the LLM that should be used for the response. 

The model adapter maintains a model pool, containing different LLMs and their attributes such as their IDs, cost-per-token, availability (e.g., different regions) and capabilities (derived from publicly available benchmarks). 
It exposes a filter based interface to select appropriate models and can \emph{combine} them based on the provided attributes and the selected \stype{}:

{
\tight
\fontsize{8.8}{10}\selectfont
\begin{verbatim}
 Filter([Model], attributes) -> [Model]
\end{verbatim}
}

Applications can specify low-level attributes such as model IDs and cost-per-token (to pick a particular model).
Alternatively, they can also choose to \emph{delegate} the choice of the LLM by using the \texttt{model\_selector} service type, letting \sys{} find the LLM best suited for the application needs.

There are many concurrent efforts to build model selection strategies~\cite{ong2024routellmlearningroutellms, ding2024hybrid, frugalgpt, llm-blender-2023, bedrock-routing}.
These can be supported as different \texttt{service\_types}.
Our implementation of delegating model selection uses a verification based strategy involving a low cost LLM ($M_1$), a high cost LLM ($M_2$) and a verifier LLM. For all queries, $M_1$ first answers the prompt. Then, the verifier judges the response on a scale of 1-10 using a pre-configured judging prompt.
$M_2$ is consulted for a final answer only if the score generated by the verifier is less than a configurable threshold.

The model adapter picks appropriate selections for these models from the model pool using suitable filters.
The heuristic it applies is that the cost-per-token of the verifier should be less than $M_1$'s which should in turn be less than $M_2$'s cost-per-token.
Applications can also specify which LLMs they desire for this strategy as key-value pairs as shown in Fig~\ref{fig:design_overview}.
In~\S\ref{sec:evaluation} we show real world benefits this simple strategy has on our production WhatsApp Q\&A dataset.

Finally, if the quality of the response is unsatisfactory, applications can invoke \texttt{proxy.regenerate()} which will directly route the prompt to the more expensive LLM.

\subsection{Context Manager}\label{sec:design_context_manager}
The context manager tracks the history of users' conversations.
This is additional input that an LLM may process along with each prompt, therefore increasing cost.
Keeping context management in the proxy 
has two key benefits: first, it allows \sys{} to optimize exactly what context is used (analogous to data compression in HTTP proxies), and secondly, it aids iterative prompt-resolution --- applications don't have to resend context each time they choose to regenerate responses.

To support several context management strategies, \sys{} uses a filter API where each filter can narrow down which messages are included in the context:

% The context manager interface must be expressive enough to support different strategies.
% To do so, 

% However, some challenges arise by limiting applications to not manipulate context in whatever way they deem necessary

% alleviates the need for applications to implement this directly, and gives \sys{} a chance to optimize exactly what context is used.

{
\tight
\fontsize{8.8}{10}\selectfont
\begin{verbatim}
 Filter([Message], prompt) -> [Message]
\end{verbatim}
}

A message is defined as a prompt-response pair.
Table~\ref{table:context} demonstrates different ways to this interface, including combining different sets of filters.

% The interface accepts a 2-dimensional array of these filters, the inner arrays combine to further filter the context, and the outer dimension joins the results from different sets of filters. This process is detailed through examples in Table~\ref{table:context}.

The default behavior is to add all available context that fits in the context window of a model.
As Fig.~\ref{subfig:quality_motivation} shows, a simple well known strategy like last-k can be much more efficient. Applications using this strategy can opt to delegate the choice of $k$ to the context manager via the \texttt{smart\_context} service type.
In this mode, the context manager uses a low-cost model (\texttt{context-LLM}) to decide how much context (i.e., value of $k$), and thus input tokens, are required. We implement this as the SmartContext filter.
% The process of using this strategy is visualized in Fig~\ref{fig:smart_context_diagram}.

% The choice of $k$ does not need to be static, it can be based on the context itself. As we saw in the previous section, we can again use an LLM for making this decision. We call this case ``SmartContext'' and implement it as another filter.

% The ``SmartContext'' filter is the primary way \sys{} reduces the number of input tokens and therefore overall cost. The LLM determining necessity of context, the \texttt{context-LLM}, must be cheaper than that used to generate the prompt reply. The process of using SmartContext is visualized in Fig

A false positive occurs when \texttt{context-LLM} wrongly excludes required context, while a false negative occurs when it wrongly includes unnecessary context.
The latter increases cost while the former reduces the quality of responses.
To reduce false positives and ensure high quality responses we invoke the \texttt{context-LLM} at most two times and only consider the prompt to not require context if both LLM calls deem it standalone.
This is feasible since \texttt{context-LLM} is relatively inexpensive and fast.

\begin{comment}
    
\begin{figure}[!t]
\begin{center}
\includegraphics[width=\columnwidth]{figures/context/ContextManager2.pdf}
\caption{SmartContext after a LastK(2) filter. First a new query is received, then the \texttt{context-LLM} processes this query and the last 2 messages. Lastly, the Chat LLM is used with or without the context to generate a response. }
\label{fig:smart_context_diagram}
\end{center}
\vspace{-1.5em}
\end{figure}
\end{comment}

\begin{table}
\small
\begin{tabular}{@{}ll@{}}
\toprule
Filters & Description \\ \midrule
% \begin{tabular}[c]{@{}l@{}}\textbf{LastK(\emph{k})}\end{tabular} &
%   \begin{tabular}[c]{@{}l@{}}Selects the most recent \emph{k}\\ messages from the context\end{tabular} \\ \midrule
\begin{tabular}[c]{@{}l@{}}\textbf{SmartContext(\emph{LLM})}\end{tabular} &
  \begin{tabular}[c]{@{}l@{}}LLM decides if context is not\\needed, otherwise all context\end{tabular} \\ \midrule
\begin{tabular}[c]{@{}l@{}}\textbf{[LastK(5), SmartContext]}\end{tabular} &
  \begin{tabular}[c]{@{}l@{}}Either the last 5 messages\\ or no context\end{tabular} \\ \midrule
\begin{tabular}[c]{@{}l@{}}\textbf{[[LastK(4), SmartContext],}\\ \textbf{LastK(1)]}\end{tabular} &
  \begin{tabular}[c]{@{}l@{}}Either the last 4 messages\\ or just the last message\end{tabular} \\ \midrule
\begin{tabular}[c]{@{}l@{}}\textbf{Similar(\emph{$\theta$})}\end{tabular} &
  \begin{tabular}[c]{@{}l@{}}Messages with similarity $>\theta$\\to the current prompt\end{tabular} \\ \midrule
\begin{tabular}[c]{@{}l@{}}\textbf{Summarize(\emph{LLM})}\end{tabular} &
\begin{tabular}[c]{@{}l@{}}LLM summarizes the context\\ messages into a single message\end{tabular} \\ \bottomrule
\end{tabular}
\caption{Examples of context API. The second example is evaluated in \S\ref{sec:microbenchmarks}. In the third example, the second dimension is used to always include one context message, even if SmartContext decides context is not necessary.}
\label{table:context}
\end{table}

The filter based API (Table~\ref{table:context}) also supports other well known context management strategies.
The ``Summarize'' filter uses the \texttt{context-LLM} to reduce a long history of messages into a short summary.
The ``Similar'' context filter returns messages in order of their similarity to the current prompt, as opposed to order of recency. This uses the vector database managed by the cache and is another reason the two components benefit from being part of the same proxy.

The design can be extended to enable richer context management strategies such as using \texttt{context-LLM} to derive users' interests, language preferences, location, upcoming events (e.g., meetings) etc.

% , and requires using the cache API to embed prompts and responses. This interplay of context management and caching

% This is a similar approach to strategies offered by other tools such as LangChain to reduce the number of input tokens so that a long conversation can fit in a small context window~\cite{langchain_summary}. 

A final consideration is how the context is updated.
Typically, when the context is retrieved it will be updated to include the next message, but this is not always the case.
Consider a chat application that has one prompt to reply to a user query and another to determine the user’s mood from past messages~\cite{twips-haroon}.
The second prompt includes the context, but does not update it.
In these cases, the coordinator must retrieve context but not insert any.
% A prefetch also has this behavior.\fahad{how?} 

\subsection{Cache}\label{sec:design_cache}

% \fahad{Will be useful to make this paragraph more crisp and shorter}
In \sys{}, we build a caching system based on the primitives offered by a vector database (i.e., semantic search).
For a semantic cache to be effective, it is important to support a rich set of operations on the \texttt{PUT} and \texttt{GET} paths.
For example, on the \texttt{PUT} path, an important consideration is the keys used to store objects (e.g., prompt vs. response) --- unlike traditional caches which typically use a single well-defined key, such as the object's hash.
% On the \texttt{GET} path, applications may want to retrieve cached objects based on different criteria (e.g., similarity).
% leverage differences in similarity scores between matched items to perform different actions (e.g., return items as is or apply common transformations to make them more suitable).

When applications desire fine-grained control, the cache interface accepts low-level specifications (e.g., similarity thresholds).
The interface also allows \sys{} to delegate responsibility to the cache, both on the \texttt{PUT} (e.g., generate appropriate keys) and \texttt{GET} (e.g., rewrite cached response) paths.
This is similar, in principle, to the delegation based strategies employed by the model adapter~(\S\ref{sec:design_model_adapter}) and context manager~(\S\ref{sec:design_context_manager}).

\paragraph{PUT operation.}
The cache needs to store objects which could be an LLM interaction (i.e., prompt-context-response trio) or an externally supplied piece of information (e.g., document).
Each object can consist of several cached types (e.g., \texttt{Prompt}, \texttt{Context} etc.) which can potentially act as keys in the database.
This is captured by the following \texttt{PUT} interface:

{
\tight
\fontsize{8.8}{10}\selectfont
\begin{verbatim}
PUT(Object, optional=[(CachedType, Key)])
\end{verbatim}
}
% The quality of the responses returned from the cache relies on strategically placing objects in the cache from the outset.
% Fundamentally, it depends on the appropriate kind of ``keys'' being used (e.g., query vs. response).

\noindent{}Embeddings --- vector representations --- are created from the keys supplied and stored in a vector database.
Generally more meaningful keys will result in more useful embeddings~\cite{hyde}.
Providing keys is optional; if they are not specified the delegated \texttt{PUT} (described later) is used.
% The behavior when no Value is provided is for the cache key to be created from the Object itself. Otherwise, a modified (e.g., summarized, redacted) version of the Object could be used as a Value.
% For example, during retrieval, the application may want to apply a stricter matching criteria with the context, and be more lenient when matching with the prompt, in scenarios where context matters a lot.
% For example, they can enable context-aware caching by including the context as a key (described later).
% In multi-user settings, cached types can enforce privacy by being created at a permission level, restricting access as needed.
% For example, a medical facility could set up a patient level cached type, ensuring the queries from one patient are not matched with medical documents for another.

\textit{Example.}
If an application wants to cache an LLM generated response with only the prompt as the key, it can specify this as:

{
\tight
% \footnotesize
\fontsize{8.8}{10}\selectfont
\begin{verbatim}
PUT('Use data structures like B-trees & Tries', 
 [(Prompt, 'How do I speed up my cache?')])
\end{verbatim}
}

A future prompt: ``Give me examples of popular data structures?'' will likely not match with ``How do I speed up my cache?'' --- the cosine similarity is 0.18\footnote{Based on OpenAI's \textsf{text-embedding-3-large}} --- but is likely to match with the response : ``Use data structures like B-trees \& Tries''
(similarity of 0.64) and can be rewritten by a small model to be more suitable for the new prompt.
Thus, the application can also use responses, and other cached types, as keys.
This can be done as follows:

{
\tight
\fontsize{8.8}{10}\selectfont
\begin{verbatim}
PUT('Use data structures like B-trees & Tries', 
 [(Prompt, 'How do I speed up my cache?'), (Response,
 'Use data structures like B-trees & Tries')])
\end{verbatim}
}

\paragraph{Delegated PUT.}
Supplying fine-grained keys hinges on the application's knowledge of future prompts and are \emph{optional} parameters of the \texttt{PUT} interface.
The delegated \texttt{PUT} mode allows applications to leave it up to the cache to decide the best key generating strategy.
This is useful when the application wants to populate the cache with complex objects (e.g., a Wikipedia article).
% with high quality information ; where the object to be cached can be long and complex.
In such settings, creating an embedding of the entire object may not be useful.
To support this delegation, the cache leverages a low-cost local model to intelligently generate keys based on the nature of the object to be cached.

In the delegate mode, the cache uses a small model (\texttt{cache-LLM}) to break down a complex object into smaller chunks and generate meaningful keys for each chunk.
In addition to using the chunk itself as the key, extra keys are generated based on: \emph{hypothetical questions} that the chunk can help answer and \emph{key-words} extracted from the chunk.
The cache also generates modified versions of the chunk: a summary and list of facts present in the chunk (useful when the workload consists of factual queries as we show in~\S\ref{sec:evaluation}).
Similar ideas have been explored by other proposals in the RAG scenario (e.g., LangChain~\cite{langchain}), motivating the benefits of making them part of our cache.

\paragraph{GET operation.}
The \texttt{GET} interface provides low-level control to applications to retrieve objects based on semantic similarity.
This is captured via a filter based API:

{
\tight
\fontsize{8.8}{10}\selectfont
\begin{verbatim}
GET([(Key, [Filter])])->[response]
\end{verbatim}
}

% Depending on the quality of match between the prompt and the cached object(s), the application may want to undertake different actions, and this is the primary way to provide control to an application over its quality vs. cost saving preferences.
% For instance, a very similar match can be returned as-is, providing the maximum cost savings, while a moderately similar one may need to be rewritten to ensure its suitability before being returned.
% This maps neatly to a match-action based interface.

Applications can provide a set of filters based on 1. cached types (e.g., \texttt{Prompt}, \texttt{Document}), 2. a minimum similarity threshold ($s$), or 3. maximum number of items to be returned ($k$).

For example, a simple look up to return all responses for which the prompt-to-prompt similarity is above a threshold (e.g., 0.9) can be specified as:

{
\tight
\fontsize{8.8}{10}\selectfont
\begin{verbatim}
GET([`How do I speed up my cache?',[(Prompt,s=0.9)]])
\end{verbatim}
}

\begin{comment}
    
\end{comment}
% \noindent{}Filters can also be \emph{combined} to facilitate uses cases where similarity requirements differ across cached types (e.g., a higher similarity threshold for context compared to query).

% {
% \tight
% \fontsize{8.8}{10}\selectfont
% \begin{verbatim}
% GET([`How do I speed up my system?',[(Query,s=0.7)],
% `I am designing a priority queue`,[(Context,s=0.9)])
% \end{verbatim}
% }

% A related example is if the application wants to match with both prompts 
% Applications can provide low-level specifications as well as rely on internal to return a high quality response.
% Depending on the similarity between the query and the cached object, the application may want to undertake different actions, and is the primary way to capture an applications quality vs. cost saving preferences.
\paragraph{Delegated GET.}
% A low-level specification relies on the application knowing the appropriate range of similarity scores or type of items to filter.
Applications can also delegate this responsibility to the cache by specifying an LLM based filter --- we call this strategy ``SmartCache''.
SmartCache internally retrieves top-$k$ items across all cached types and determines whether the retrieved objects are relevant/appropriate (similar to SmartContext~\S\ref{sec:design_context_manager}).
It then uses the retrieved objects to generate a suitable response.
The response could be 1. the cached object as-is, 2. a rewritten response or 3. one generated using the user's prompt, context and the cached information.

\section{Implementation}
\label{sec:implementation}
\sys{} has been in production for over a year.
It is implemented as a Python application running in WAS Lambda functions~\cite{aws}.
It offers access to 10+ LLMs, including OpenAI~\cite{openai_assistant}, Anthropic~\cite{anthropic}, Llama~\cite{llama}, and Phi~\cite{phi} models.
The necessary state such as the conversation history is stored in key-value stores (DynamoDB) and a SQL table (RDS) with vector-search is used to support the cache.

Each incoming request is first converted into an embedding (we use OpenAI embeddings~\cite{openai_embeddings}) and looked up in the cache (\S{\ref{sec:design_cache}}) for a response.
If the cached response is not available or used, the context manager (\S\ref{sec:design_context_manager}) retrieves past messages from the conversation history.
Finally, the model adapter (\S{\ref{sec:design_model_adapter}}) is queried to use provider-specific APIs to generate an LLM response.
% Before \sys{} returns a response, the new query/response pair is optionally added to the NoSQL table.

LLMs can vary widely in time to fully generate a response, particularly when we are combining models.
To ensure requests are processed in the expected order we use a per-user FIFO queue (AWS SQS).
Every incoming request goes through this queue, and is only removed from the queue when a response has been sent.

Using a serverless architecture has made it convenient to set up a development and production environment --- a useful enabler for incrementally adding features.
Production is a stable copy of the various functions and continues serving user requests.
Another key benefit is reduced cost; since the functions themselves are light-weight (in compute and memory requirements), they are amenable to the serverless architecture.
To reduce the impact of cold starts, which can be >1s according to our measurements, all features of the proxy are in one serverless function.

\begin{figure}[h!]
    \centering
    \begin{subfigure}[t]{0.47\columnwidth}
        \centering
        \adjustbox{valign=t}{\includegraphics[width=\textwidth]{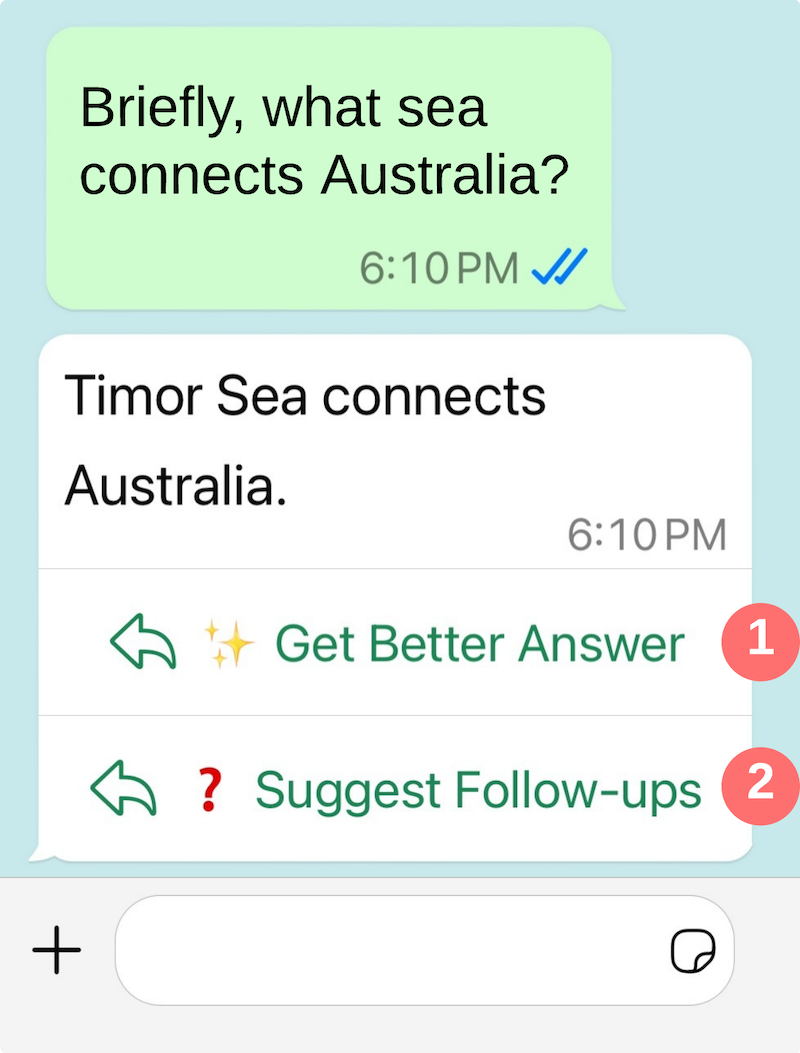}}
        \caption{User query and response.}
        \label{subfig:wa-layout}
    \end{subfigure}
    \hfill
    \begin{subfigure}[t]{0.47\columnwidth}
        \centering
        \adjustbox{valign=t}{\includegraphics[width=\textwidth]{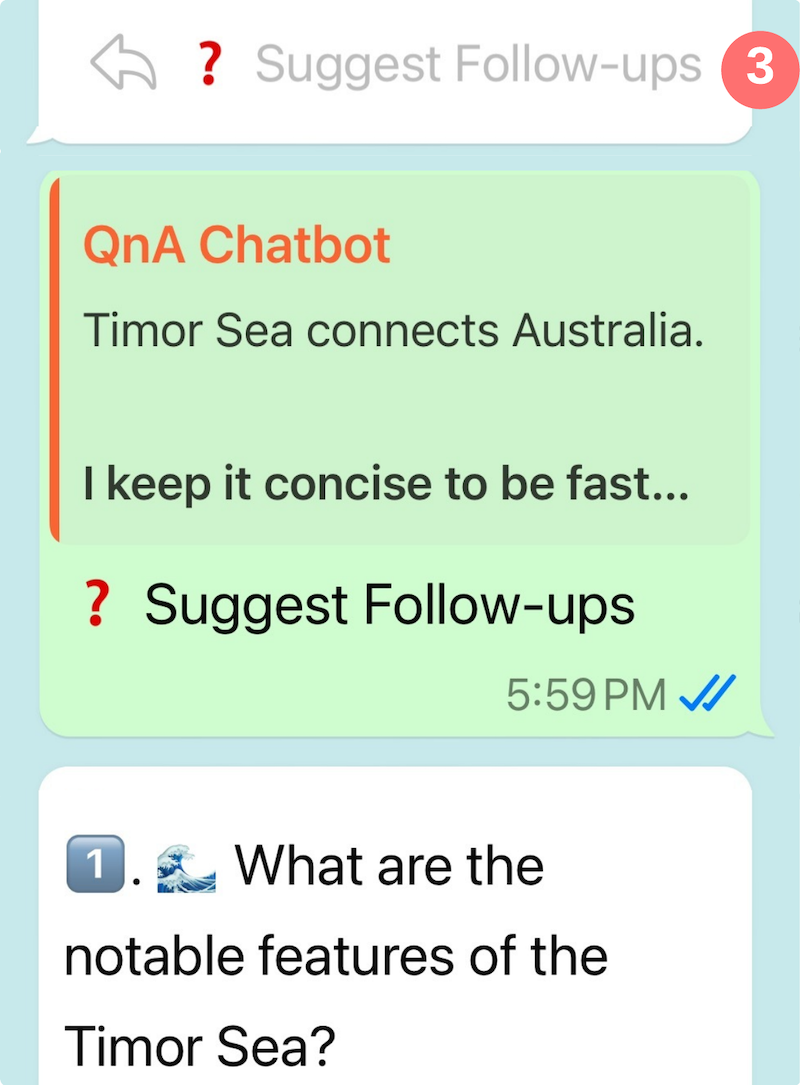}}
        \caption{Response to button 3.}
        \label{subfig:wa-follow}
    \end{subfigure}
    % \hfill
    % \begin{subfigure}[t]{0.3\columnwidth}
    %     \centering
    %     \adjustbox{valign=t}{\includegraphics[width=\textwidth]{figures/whatsapp/wa-moment.png}}
    %     \caption{Message sent to user to mask latency, as pre-fetch was not complete before click.}
    %     \label{subfig:wa-moment}
    % \end{subfigure}
    \caption{The WhatsApp Q\&A service. Buttons 1-3 have pre-fetched (and cached) responses which are returned when a user interacts with them to avoid delays and keep the conversation responsive.
    % \textcolor{red}{This figure is not referenced in the text.}
    }
    \label{fig:wa-screenshots}
\end{figure}

%\include{case-study}

% \section{Evaluation}\label{sec:evaluation}
% In this section, we first explore the diverse range of services built on top of \sys{}.
% Next, we share our insights from developing and deploying a WhatsApp-based Q\&A service~(\S\ref{sec:deployment}).
% Finally, we present microbenchmarks of various components of \sys{}~(\S\ref{sec:microbenchmarks}).

\section{Evaluation}\label{sec:evaluation}
To understand \sys{}'s generalizability and effectiveness across real-world workloads, we evaluate its use in a range of settings that vary in user expertise, latency sensitivity, and cost constraints.
We begin by highlighting the diverse range of applications built using \sys{}.

We then present a detailed case study of our WhatsApp-based Q\&A service~(\S\ref{sec:deployment}), showcasing the behavior and impact in a real-world, cost- and latency-sensitive deployment.
Next, we describe its use in academic course projects~(\S\ref{sec:course-projects}), which offers insight into \sys{}'s accessibility, flexibility, and usability among student developers.
Finally, we present microbenchmarks of \sys{}’s core components~(\S\ref{sec:microbenchmarks}), quantifying the cost, latency, and quality trade-offs enabled by its design.

% \usepackage{booktabs}

\begin{table}[]
\resizebox{\columnwidth}{!}{%
\begin{tabular}{@{}ll@{}}
\toprule
\multicolumn{1}{c}{Prototypes \& Projects}                                         & Features                                                                   \\ \midrule
\begin{tabular}[c]{@{}l@{}}\textbf{TWIPS}: LLM powered texting\\ app for Autistic users~\cite{twips-haroon}\end{tabular} &
  \begin{tabular}[c]{@{}l@{}}Determine user tone (simple)\\ Improve user messages (complex)\end{tabular} \\ \midrule
\begin{tabular}[c]{@{}l@{}}\textbf{Morshid}: LLM powered\\ educational app\end{tabular}     & \begin{tabular}[c]{@{}l@{}}Similar questions\\ asked by users\end{tabular} \\ \midrule
\begin{tabular}[c]{@{}l@{}}\textbf{LLM-based HTTP proxies}: LLMs\\ for improving webpage accessibility\end{tabular} & \begin{tabular}[c]{@{}l@{}}Large HTML pages\\ included as context\end{tabular} \\ \bottomrule
\end{tabular}%
}
\caption{Use cases of \sys{} that supports features required by various applications}
\label{tab:use-cases}
\end{table}

% \begin{table}[]
% \resizebox{\columnwidth}{!}{%
% \begin{tabular}{@{}cccc@{}}
% \toprule
%          & Testbed (16 GPUs) & \multicolumn{2}{c}{Simulations (64 GPUs)} \\ \midrule
% Workload & Workload-1        & Workload-2          & Workload-3          \\
% Job Type & AutoML            & DNN                 & DNN                 \\
% DNN/job  & 1-20              & 1                   & 1                   \\
% GPUs/DNN & 1                 & 1-52                 & 1-8                \\ \bottomrule
% \end{tabular}%
% }
% \caption{Summary of the settings used to evaluate \sys{}}
% \label{table:eval-table}
% \end{table}

% \paragraph{Supporting different applications.}
% \sys{} has been used to build a variety of applications, including research prototypes, academic course projects, and a WhatsApp-based Q\&A service that we developed.
% Research prototypes include AI-powered assistive technologies and educational services, whereas course projects have focused on designing web accessibility features, chatbots for social using our proxy.
% Table~\ref{tab:use-cases} summarizes these and highlights features which have benefited from model selection, caching, and context management.

\paragraph{Supporting different applications.}
\sys{} has been used to build a variety of applications, including research prototypes, academic course projects, and a WhatsApp-based Q\&A service that we developed.
Research prototypes include AI-powered assistive technologies (e.g., detecting user tone or rephrasing messages) and educational tools.
Academic course projects have focused on AI powered web accessibility enhancement, multi-agent systems, and chatbots addressing social good applications.
Across these use cases, \sys{}’s unified interface and built-in support for context management, caching, and model selection enabled rapid prototyping and experimentation, even by novice users.
Table~\ref{tab:use-cases} summarizes some of these applications and highlights some of their specific features that benefit from \sys{}. \hiba{the use cases table doesn't include the multi-agent, and social good chatbots but we talked abt them in the text}

A unifying theme across these settings is cost-sensitivity: research prototypes and course projects were developed by students operating under budget constraints.
Similarly, the user base of the Q\&A service, described in the next section, \emph{primarily} comprises individuals from developing regions (e.g., Pakistan, Sudan) and members of the diaspora in the United States — all of whom are cost-conscious.

% Next, we detail our flagship service and share key deployment insights.
\subsection{Case Study I: WhatsApp Q\&A Service}\label{sec:deployment}
We have built and deployed a WhatsApp based Q\&A service using \sys{}.
Our small-scale deployment has been rolled out for over 12 months, during which over 100 users, across different countries (e.g., Pakistan, Sudan, UAE)\hiba{I would remove UAE since it is not mentioned in our IRB}, have subscribed and sent over 14.7K requests (4K free-form messages)\hiba{we can change free-form to user-generated for carity}.
We share the service's rich set of features, challenges unique to a WhatsApp based deployment (e.g., message oriented nature) and how the proxy helps to support these features.\footnote{A separate paper describes this service in detail. Here we focus on the its usage of \sys{} features.}

The Q\&A service provides its users access to the latest LLMs via WhatsApp's familiar interface (Fig.~\ref{fig:wa-screenshots}).
Cost considerations are crucial since a sizeable fraction of our user base is from developing regions where WhatsApp is popular~\cite{whatsapp-popularity}. 
At a basic level, users type-in and send their queries (topics range from health to politics and sports) to our service and get a response.
To provide a good user experience, our service supports a number of features: i) anticipating follow-up queries and pre-fetching (and caching) suitable content to enhance responsiveness; follow-up queries show up at the end of the response as buttons, ii) allowing users to regenerate a response, typically more detailed using a higher quality model, iii) pushing recommended content (e.g., trending questions, recent questions, etc.) to users, iv) giving points to users on asking questions and maintaining a leaderboard with daily and overall rankings.

These features also use the limited but powerful WhatsApp affordances (e.g., buttons), and have pushed-based content (e.g., question of the day, questions asked by others) which nudge users to opt for options that are already cached --- 13\% of the total interactions consist of users requesting the cached content.
These features have led to a notable user engagement level, with 20\% of users active for several (>10) days, and at least 300 requests sent to our service per week across users.
We next discuss how the deployment used various aspects of \sys{}.

%When a user's query arrives to the service, it launches several %tasks in addition to generating a response~(\ref{fig:whatsapp-%workflow}).
%These include anticipating follow-up queries and pre-fetching %suitable content to enhance responsiveness, pushing recommended %content (e.g., trending questions) to users to maintain engagement %%(20\% of our user base has been active for several (>10) days), %and allowing users to regenerate responses.

% causing some latency inflation due to function cold starts.

% Latency masking involves generating a shorter message , and sending to the user ASAP while prefetching a longer response.
% During thinking/reading time, we also prefetch follow-ups and suggest them to the user.
% The prefetched content is cached and later retrieved using exact match, if the user is interested.

\textit{Model Adapter.}
Having a unified interface to access different LLMs has offered ease of use.
Our Q\&A service richly leverages the capabilities of various AI models; within (GPT4o vs GPT4o-mini) and across (OpenAI vs Anthropic) LLM-families.
Tasks range from responding to user queries, generating user interests, and identifying queries with broad appeal to generate recommended content for our user-base.
These tasks have also required combining models; for example using a cheap LLM (Haiku) to filter out candidates from a large set of queries and judiciously applying an expensive model (GPT4o) to identify those likely to be popular.
Additionally, with newer and improved LLMs, we have shifted from using GPT-3.5 to GPT4o-mini without a significant difference in the quality of responses sent to our users.
While user queries have largely remained the same in complexity, newer model families have become smarter, narrowing the quality gap between cheaper models and more expensive ones --- effectively delivering more intelligence at a lower cost.
% While there was once a clear quality gap between cheaper models like GPT-3.5 and more expensive ones like GPT-4, this gap has narrowed with newer model families, likely due to the combination of enhanced capabilities of newer LLMs and the complexity of user queries remaining the same.

Different models exhibit varying latency characteristics.
% Our deployment logs show that LLM latencies can range from a few seconds (0.2s) to several (75s).
For example, our deployment logs show for larger models (e.g., GPT4o, GPT3.5) the mean (p99.9) latency is $3.8s$ ($78s$) while for smaller ones (e.g., Haiku, GPT4o-mini) it is $1.2s$ ($15s$).
These characteristics have motivated us to experiment with ``latency-centric'' \stypes{} as well.
For example, the Q\&A service uses the fastest (and also cheap) model to generate a short initial response (achieved via a suitable prompt) to a query while pre-fetching a higher quality response asynchronously from a more expensive model. This can be elicited via a ``Get Better Answer'' button (Fig.\ref{subfig:wa-layout}).
% and relies on the user asking for a regenerate which triggers a call to a higher quality model.

% We are also experimenting with displaying a wait-message to the users as soon as their message comes.

\textit{Context Management.}
Having a context management module in the proxy facilitates seamlessly switching between different models, and more importantly across different family of models, \emph{during} a conversation.
For our service, the context manager maintains user messages in chronological order and manages a few nuances including the scenario where a user requests a regeneration of their response, in which case the initial response is removed from the context.
% Noah: I commented this out because it seemed to raise more questions about the implementation than it clarified.
% For this, a special message is added in the user's context to denote it was a regenerate request.

% By decoupling the context from the models, we have observed ``in-context'' learning:
% previous responses generated by a model, passed as context to a different model, influence its response. This has both positive and negative implications. 
% On the positive side, we observed that lower quality models start providing better responses when they have examples of higher quality models as part of their context. 
% Similarly, models start to inherit other linguistic styles (e.g., tone of another model).
% These differences also lead to inconsistencies.
% For example, using models with real-time grounding capabilities (e.g., gemini-flash-2.0~\cite{gemini_2_0_flash}) that cite the sources (e.g., URLs) used, for earlier conversations can lead to other models (without grounding capabilities), hallucinating sources for responses generated afterwards.
% Future work could explore ways to bridge those differences to provide a consistent user experience by designing improved context management strategies. 

By decoupling context from specific models, we have observed a form of in-context learning: responses generated by one model, when passed as context to a different model, can influence that model's behavior.
This has both positive and negative implications.
On the positive side, lower-quality models often begin to produce improved outputs when their context includes messages generated from higher-quality models.
Similarly, models may adopt stylistic elements --- such as tone --- from the responses of other models.
However, these inherited behaviors can also introduce inconsistencies.
For instance, when responses generated using a grounded model (e.g., Gemini 2.0 Flash~\cite{gemini_2_0_flash}) --- which include citations like URLs --- are used as context for a model without grounding capabilities, the latter may hallucinate sources in its responses.
Future work could explore context management strategies to bridge these differences and ensure a more consistent user experience.

\textit{Caching.}
LLM applications often employ streaming to hide the end-to-end latency of generating a response. However, WhatsApp is message oriented, requiring creative ways to mask latency.
%The proxy can bridge the gap between streaming based generators and a message %oriented consumer, such as the Q\&A service.
Our service aggressively pre-fetches data and uses the cache as a masking strategy.
%Additionally, the cache becomes useful here.
Specifically, the Q\&A service anticipates follow-up queries the user may have.
These are generated using an LLM and stored in the cache, and are explicitly suggested as buttons (Fig.~\ref{subfig:wa-follow}).
The proxy uses an exact match to retrieve them, in case the user presses the buttons.
This is in addition to using the cache for semantic matches, which we evaluate later. 
% \fahad{forward pointer to the cache eval}
\abd{add latency numbers - when cached, how many, etc.}
%These can be leveraged to enhance the prefetching technique (e.g., selective %prefetching).

% We mask latency of LLMs by sending an immediate reply to user messages that confirm the query was received.
% Although \sys{} does not support streaming responses, streaming APIs are used internally to allow logging of the time to first token.
% Our implementation runs on AWS using Lambda functions, API Gateway, DynamoDB, Simple Queue Service, and Relational Database Service~\cite{aws}.

% and made it straightforward to test changes in a development environment that was a copy of every serverless function while production users were still served with stable code.

% Approximately 20\% of these users were active only for one day, while another 20\% have been active users for over 10 days.
% The application has features other than just question/response, such as suggested and trending questions.
% The application has received over 7K events handling interactions with these features.

% The proxy is deployed along with a WhatsApp chatbot which has been serving production users since for over 15 weeks.
% During this time over 100 users registered and sent over 1.3k messages.

\subsection{Case Study II: LLMs for Classroom Settings}\label{sec:course-projects}

Following our deployment in a production-facing Q\&A service, we evaluated \sys{} in a second cost-sensitive setting: undergraduate and graduate classrooms.
A subset of \sys{}'s features was exposed via a RESTful API to approximately 60 students across three computer science courses, where it was used to build a variety of LLM-powered applications — including web accessibility enhancement features, multi-agent reasoning systems, and chatbots for social good.
Over the course of 145 days, students issued approximately 75K requests, averaging $\approx$500 requests per day.
Several projects spanned the full semester, demonstrating \sys{}'s ability to support sustained, iterative development and maintain responsiveness under long-running, educational workloads.

Although students were not directly charged for LLM usage, this deployment reflects the practical constraints of instructional settings, where instructors or institutions typically bear infrastructure costs.
Compounding this, students came in with widely varying levels of programming and LLM experience, which meant that the system had to be both cost-efficient and approachable for novice users.
This combination of cost oversight and user diversity introduced a distinct but equally challenging deployment axis compared to the WhatsApp service.
In the face of these design pressures, 
\sys{} offered a low barrier to entry, enabling quick iteration, and experimentation while abstracting away provider-specific complexity and cost-related pitfalls. 
% while still exposing advanced features like model selection and cache-backed content reuse, enabling a broad range of projects under tight resource and time constraints.

% \sys{}'s high-level API 
% The system 

\textit{Usage-based \stypes{}.}
To keep costs predictable, we implemented a usage-based \texttt{service\_type} for \sys{}.
This allowed us to limit access to a curated set of relatively inexpensive models --- GPT4o-mini, Azure Phi-3, Claude Haiku, and Meta LLaMA-3 variants --- similar in spirit to domain denylists used in HTTP proxies.
The system also supported usage quotas based on input/output tokens and request counts.
Despite these limits, students successfully built rich, domain-specific applications.

\textit{Supporting RAG-style workflows.}
Students leveraged \sys{}'s context management and caching components to implement retrieval-augmented workflows.
They uploaded diverse reference materials — including policy documents, FAQs, and course-specific content — which the \texttt{cache-LLM} automatically chunked and indexed for semantic retrieval.
A key challenge was the structural variability of these documents: policy files benefited from section-based chunking, while FAQs required segmentation around question–answer pairs, and so on.
\sys{}'s delegated caching interface handled these differences, allowing students to store rich context and later retrieve semantically relevant chunks using a proximity-based threshold.
This ability to reuse cached content across prompts as context, allowed us to keep total LLM inference costs under \$10 across all three courses.

\textit{Models used.}
Usage logs showed that 73\% of all requests were directed to GPT4o-mini, followed by 13\% each for Claude Haiku and Meta LLaMA-3 variants, and 1\% for Phi-3.
Students who experimented with multiple models typically did so to benchmark response quality for their specific workloads.

This usage pattern has motivated us to introduce a batch-mode interface in the future, where users can submit a batch of prompts to be processed by multiple models simultaneously.
Such a feature would lower the development overhead of benchmarking and compositional workflows, while aligning with pedagogical goals, such as teaching students to reason about the importance of prompts, and trade-offs in cost, latency, and output quality across models~\cite{chainforge}.

Another interesting observation --- albeit from a single student's project in the multi-agent reasoning systems course --- was a connection between the models used and the nature of the prompts they received\footnote{A chi-squared test over the prompt distribution confirmed a statistically significant association between prompt type and model used (p < 0.001).}.
This student employed three models: Phi-3, GPT4o-mini, and Claude Haiku.
A closer inspection of their logs revealed that prompts sent to Phi-3 were generally more structured, rule-based, and imperative in tone, and often used simple, command-style grammar.
In contrast, prompts directed to GPT4o-mini and Claude Haiku tended to be less rigid, incorporating softer constraints and exhibiting a more collaborative or conversational tone.
These stylistic and structural differences may reflect the student's perception of each model’s strengths, with Phi-3 being favored for precise, deterministic behavior and the others for more nuanced or flexible responses.
Future versions of the API could surface such model-specific characteristics as part of higher-level \stypes{}.

Qualitative feedback reinforced these findings.
In post-course surveys,\footnote{The survey was distributed to a subset of students; we report results based on completed responses.} 75\% of respondents indicated that \sys{} was easy to get started with, and over half said it integrated smoothly into their project workflows.
Students particularly appreciated the simplicity of switching between models and the ability to reuse cached content across queries; features that were especially helpful for minimizing costs and supporting those with limited prior experience building LLM-based applications.

Together, these deployments show that \sys{} supports effective experimentation and development in resource-constrained settings, from cost-sensitive production services to educational environments with diverse user expertise.
We now leverage the real-world workloads gathered from the WhatsApp deployment to benchmark different cost-optimization strategies supported by \sys{}. We highlight their trade-offs in terms of cost, latency, and output quality.

\subsection{Microbenchmarks}\label{sec:microbenchmarks}

We present the results of the optimization strategies in each component of \sys{}.
While each strategy is similar to those proposed in related work~(\S\ref{sec:related-work}), our evaluation uses a production dataset (\emph{D}) of 10 conversations selected from one month of our production WhatsApp Q\&A service with > 10 messages in each conversation. In total there are 244 queries.

\paragraph{Model Selection.}
We evaluate the verification-based model selection strategy discussed in~\S\ref{sec:design_model_adapter}, which shows how to intelligently combine a cheaper and expensive model to save costs with little impact on the quality of responses.

\begin{figure}[!t]
    \centering
    \subfloat[][Original Models]{\includegraphics[width=0.48\columnwidth]{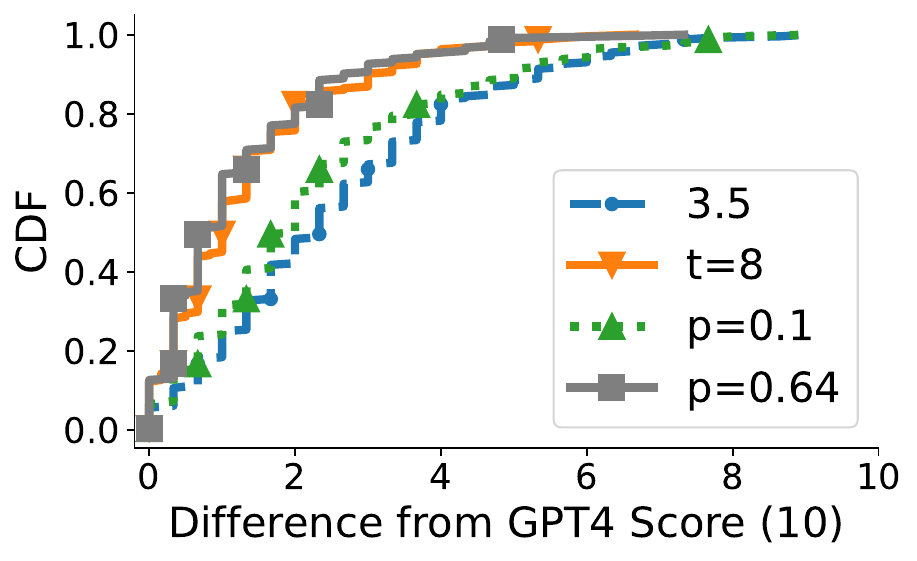}
    \label{subfig:ms_old_models}}
    \subfloat[][New Models]{\includegraphics[width=0.48\columnwidth]{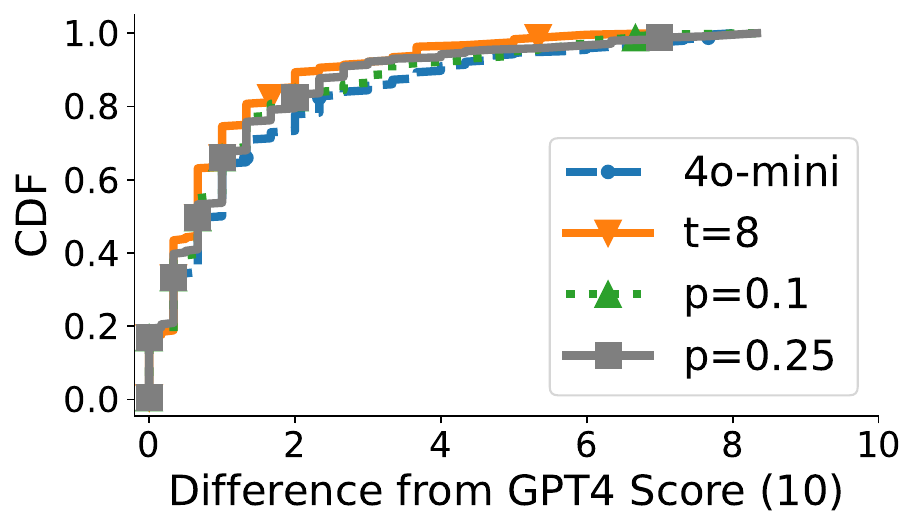}
    \label{subfig:ms_new_models}}

    \caption{Fig.~\ref{subfig:ms_old_models} compares the quality of verification with $t=8$ and random strategies with $p=0.64$, $p=0.1$ using an earlier generation of models (GPT 3.5, GPT4, Opus). Fig.~\ref{subfig:ms_new_models} is the same but with new models (GPT4o-mini, GPT4o).}
    \label{fig:opus_quality_comparisons}
    % \vspace{-1.5em}
\end{figure}

\textit{Setup.}
We evaluate using both models that were available at the time we collected the data as well newer models that are the latest at the time of writing. First, we set the less expensive model ($M_1$) to GPT3.5, expensive model ($M_2$) to GPT4, and use Claude Opus as our verifier.
The newer models are GPT4o-mini as $M_1$ and GPT4o as $M_2$ and the verifier.
Both are evaluated on \emph{D} using the strategy described in~\S\ref{sec:design_model_adapter}.
This is compared to only using $M_1$ to answer all the questions.
The response from $M_2$ is assumed as the reference, and hence always gets a score of 10.
We compare our intelligent strategy with a strategy that randomly selects a model --- a common practice in optimization (e.g., hyperparameter tuning~\cite{he2021automl}).
This strategy randomly uses $M_2$ with a probability of $p$, and otherwise uses $M_1$.
With $t=8$, $M_2$ is used to answer > 60\% of the prompts with the original models and 25\% of the prompts with new models. Based on this, each experiment shows the random strategy $p=0.64$ or $p=0.25$ as well as $p=0.1$ to demonstrate a lower cost alternative.

\textit{Results.} As shown in Fig.~\ref{subfig:ms_old_models}, our verification strategy noticeably outperforms using $M_1$ all the time and has noticeably more answers within 1 to 3 pts of $M_2$'s answers than $M_1$.
Interestingly, Fig.~\ref{subfig:ms_new_models} indicates that newer generation of models are capable of answering the kinds of questions users ask our service even with the cheaper variants (4o-mini).
This result is also demonstrated by the proportions of prompts that are routed to each model. When the 4o family of models is used a smaller percentage of prompts are routed to the large model.
While the $p=0.64$ random strategy performs similarly to our model selection, picking the optimal percentage a priori can be challenging ($p=0.1$ is worse).
% We also expect there would be a greater difference if a smaller percentage of queries routed to $m2$ because the random strategy would more often use $M_2$ for less complex queries.

\begin{figure}[!t]
    \centering
    \subfloat[][Cost Comparison]{\includegraphics[width=0.48\columnwidth]{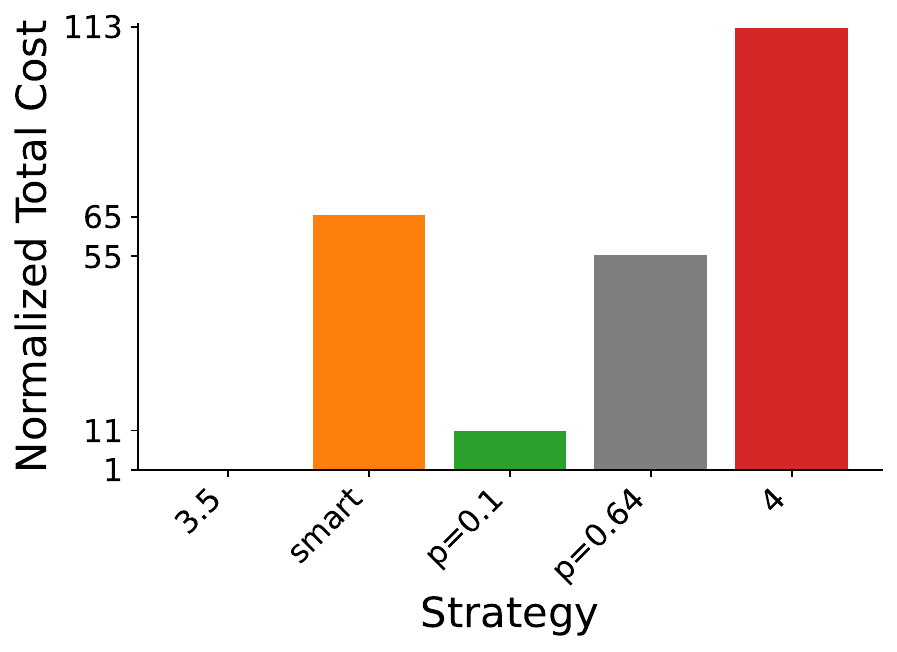}
    \label{subfig:opus_v_eq_quality_cost}}
    \subfloat[][Time Comparison]{\includegraphics[width=0.48\columnwidth]{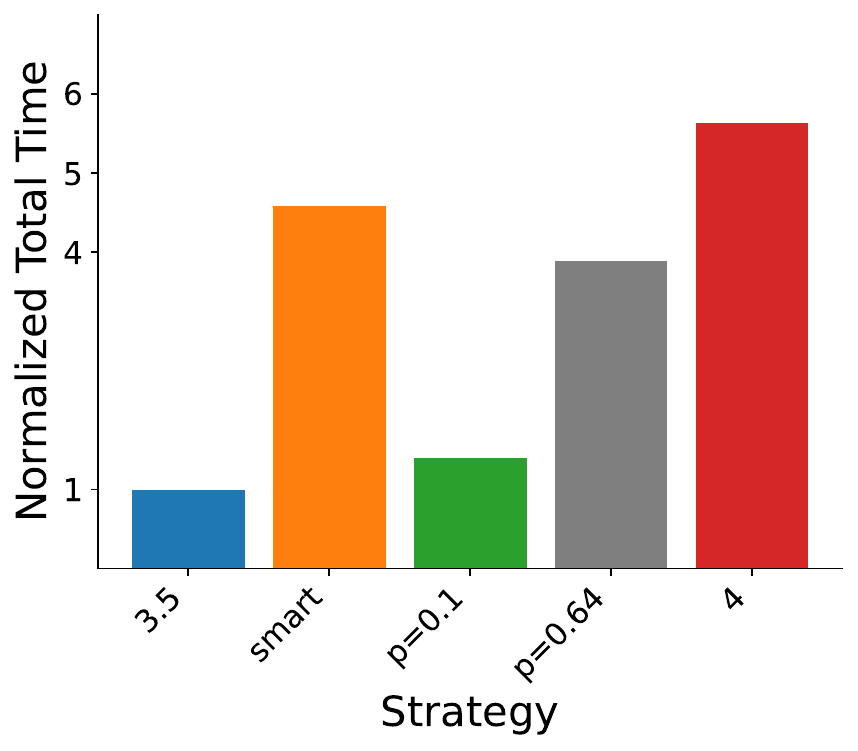}
    \label{subfig:opus_v_eq_quality_time}}

    \caption{\ref{subfig:opus_v_eq_quality_cost} compares the cost of answering all prompts using our verification strategy with $t=8$ and our random strategy with $p=0.64$. \ref{subfig:opus_v_eq_quality_time} compares the total time. Both are normalized to GPT3.5}
    \label{fig:opus_perf_comparisons}
\end{figure}

In Fig.~\ref{fig:opus_perf_comparisons} we show the cost and time comparisons of these strategies using older generation models.
Fig.~\ref{subfig:opus_v_eq_quality_cost} shows the (normalized) total cost of each corresponding strategy and demonstrates that our model selection has a 40\% reduction in cost from using $M_2$ only.
Fig.~\ref{subfig:opus_v_eq_quality_time} demonstrates that model selection is significantly faster than using $M_2$ exclusively, although it is about 5\ttimes{} slower than using only $M_1$.
Overall, these results show that our simple model selection can reduce costs while maintaining quality.
Consistent with these results, we updated
our Q\&A service to use GPT4o-mini instead of GPT3.5.

% \subsection{}\label{sec:evaluation_context_manager}
\paragraph{Context Manager.}
We evaluate the \texttt{smart\_context} service type, which uses a low cost LLM to decide if context needs to be supplied to a high cost LLM.
We see an up to 50\% reduction in cost compared to last-k, while limiting the tail of low quality responses resulting from using no context~(\S\ref{sec:motivation_context}).

\textit{Setup.}
Queries in \emph{D} were replayed with following strategies: last-k with k=0 (no context), 1 (most recent message), 5 (baseline), SmartContext with k=1, and SmartContext with k=5.
% \begin{description}[noitemsep]
% \item [LastK(5):] Include the last five context messages. This is our baseline and considered the highest quality response.
% \item [LastK(1):] Including only the last context message.
% \item [LastK(0):] Each prompt evaluated with no context.
% \item [{[Lastk(5), SmartContext(GPT-4o mini)]:}] A smaller LLM decides if the last five context messages are needed. If not, the prompt gets no context.
% \item [{[Lastk(1), SmartContext(GPT-4o mini)]:}] A smaller LLM decides if the last context message is needed. If not, the prompt gets no context.
% \end{description}
After replaying each conversation we judged the quality of conversations with the LastK(5) conversation used as reference.
% We manually spot-checked and observed this to be enough context in our sample conversations, making it a reasonable choice for a high-quality baseline despite not necessarily the highest possible quality.

For each of the $N$ messages in a conversation, $C_{i}$, and the reference conversation, $R_{i}$, where $0 \leq{} i \leq{} N$, the judge gave a score $0 \leq{} S_{i} \leq{} 10$ based on inputs $C_{i}$, $C_{i-1}$, $R_{i}$, $R_{i-1}$.
% One previous message is used to provide the judge some context but not so much that one bad response greatly affects the remaining conversation.
The results are averaged across three runs.
% Due to inherent randomness in the LLM APIs we ran each strategy three times, the resulting scores and costs are averaged.

\begin{figure*}
    \centering
    \subfloat[][Cost]{\includegraphics[width=0.25\textwidth]{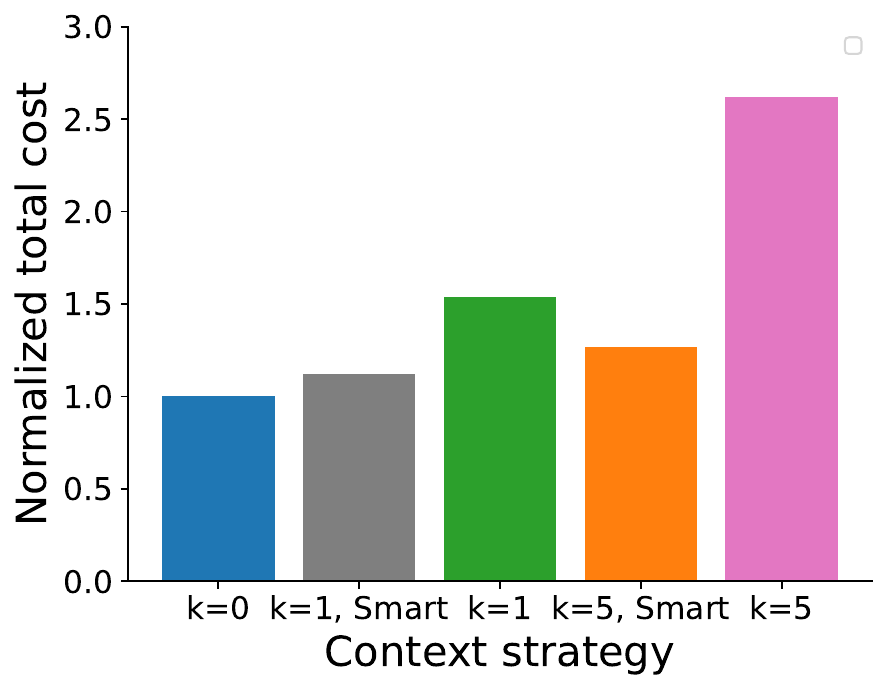}
    \label{subfig:context_cost}}
    \subfloat[][Quality]{\includegraphics[width=0.25\textwidth]{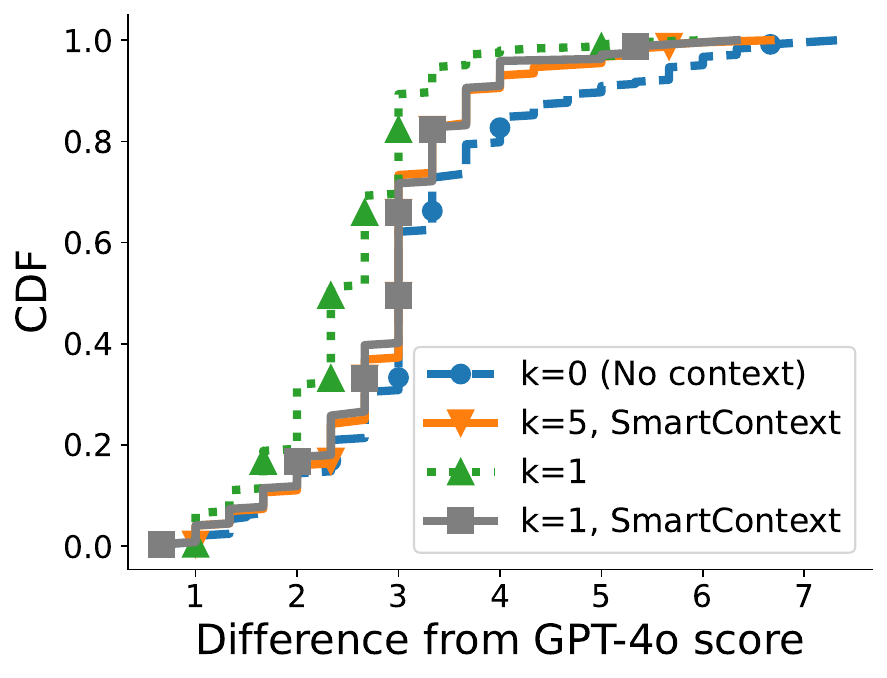}
    \label{subfig:context_quality}}
    \subfloat[][Time]{\includegraphics[width=0.25\textwidth]{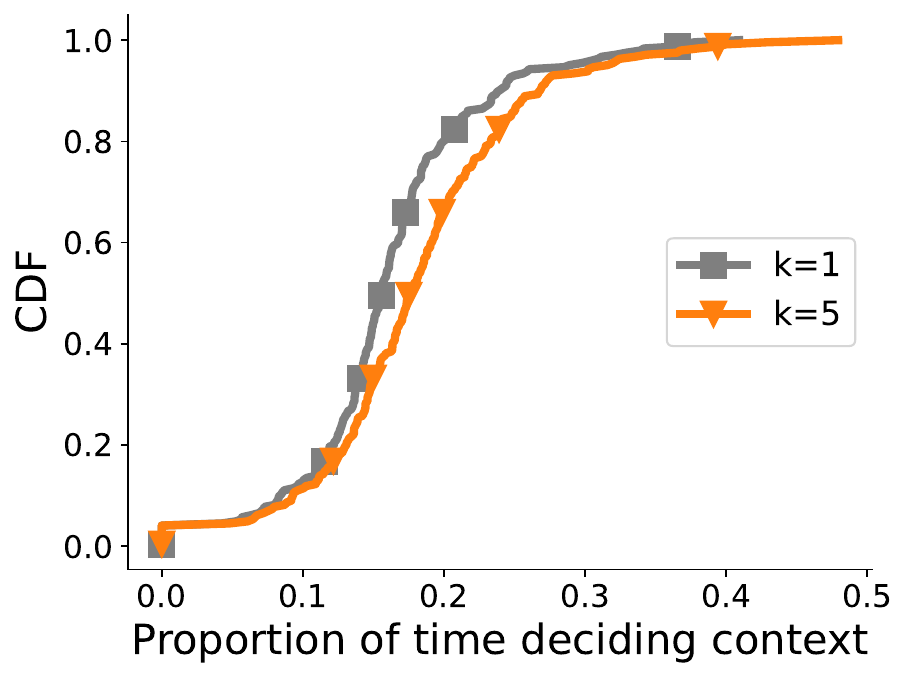}
    \label{subfig:context_time}}

    \caption{Results of context experiments. \ref{subfig:context_cost} shows cost, normalized with the lowest to 1, for each strategy. No context is cheapest, as expected. Smart strategies are $\sim$30\% and $\sim$50\% cheaper for k=1 and k=5, respectively. \ref{subfig:context_quality} is a CDF of response quality for each strategy. k=0 has the worse quality, as expected. Both smart context strategies are similar in quality, falling between k=0 and k=1. k=5 is the baseline that quality is scored against. \ref{subfig:context_time} is a CDF of the proportion of time replying to each prompt that is spent determining if context should be used for the k=1 and k=5 SmartContext strategy.}
    \label{fig:context_experiment_results}
\end{figure*}

\textit{Results.} The results of our experiments for cost, quality, and time are shown in Fig.~\ref{fig:context_experiment_results}.
Results show SmartContext combined with $k=1$ or $k=5$ can reduce costs by 30-50\% while being higher quality than the no-context response. Quality is similar whether $k=1$ or $k=5$, suggesting most of the quality difference improvement is present with just one message in the context.
SmartContext quality is particularly higher than no-context in the tail 20\% of queries, following from our intuition that only some queries require the context.
This shows how SmartContext only slightly reduces quality for large benefits in cost-savings.

The extra LLM call is likely to increase total time to generate a response. We keep the number of output tokens of the intermediate LLM call small, and our results indicate the added time using the SmartContext strategy accounts for < 20\% of total time for about 80\% of messages when k=1. The largest time used by this extra LLM call is less than 50\% of the overall request time.

\paragraph{Cache.}
We evaluate the \texttt{smart\_cache} service type, which uses a local small model combined with high quality cached information to generate responses.
An issue with small models is their tendency to hallucinate (especially with factual information).
% \fahad{State in a line or two the high level setup of this experiment -- that it is going to use high quality stuff in the cache with a low cost LLM  to answer new queries}
For such queries, \texttt{smart\_cache} is able to improve the worst-case quality by 4\ttimes{}.
While similar to RAG systems~\cite{RAG}, an interesting insight is the use of widely available information sources to \emph{intelligently} populate the cache.

\textit{Setup.}
The cache is populated with Wikipedia~\cite{wikipedia} articles on topics gathered from our WhatsApp service usage, using the delegated \texttt{PUT}.
% The cache breaks them down into smaller chunks.
% Chunks serve as keys, with the corresponding response representing a list of facts extracted from the chunk.
% \fahad{same comment as earlier -- is there a method/criteria for picking these conversations. If yes, state that}
We select 170 queries across 17 user conversations.
These queries represent the last 10 requests per user, at the time of running the experiment, sent to the Q\&A service.

We focus on queries that are factual (using GPT4o to determine this), which consist of $~30\%$ of the overall queries (i.e., 51 queries), since this has the largest opportunity in leveraging a cache populated with factual information.

The \texttt{smart\_cache} uses Phi-3~\cite{phi3} (3.8B parameter model).
We compare our approach against directly (i.e., with cache disabled) using \underline{GPT4o} and \underline{Phi-3} to answer queries.

Conversations are replayed and the response quality is judged with a reference answer generated using Sonar-Huge-Online~\cite{perplexity}.
% Responses are generated and judged three times and we compute the average score for each strategy (out of 10).
% The reference answer is generated by Perplexity using their online model: 
% \footnote{\texttt{llama-3.1-sonar-huge-128k-online}~\cite{perplexity}}.
Sonar-Huge-Online serves as a strong baseline since it has access to information on the internet and the responses it generates are factually grounded; an important facet of our experiment.

\begin{figure}[!t]
    \centering
    \subfloat[][Overall]{\includegraphics[width=0.48\columnwidth]{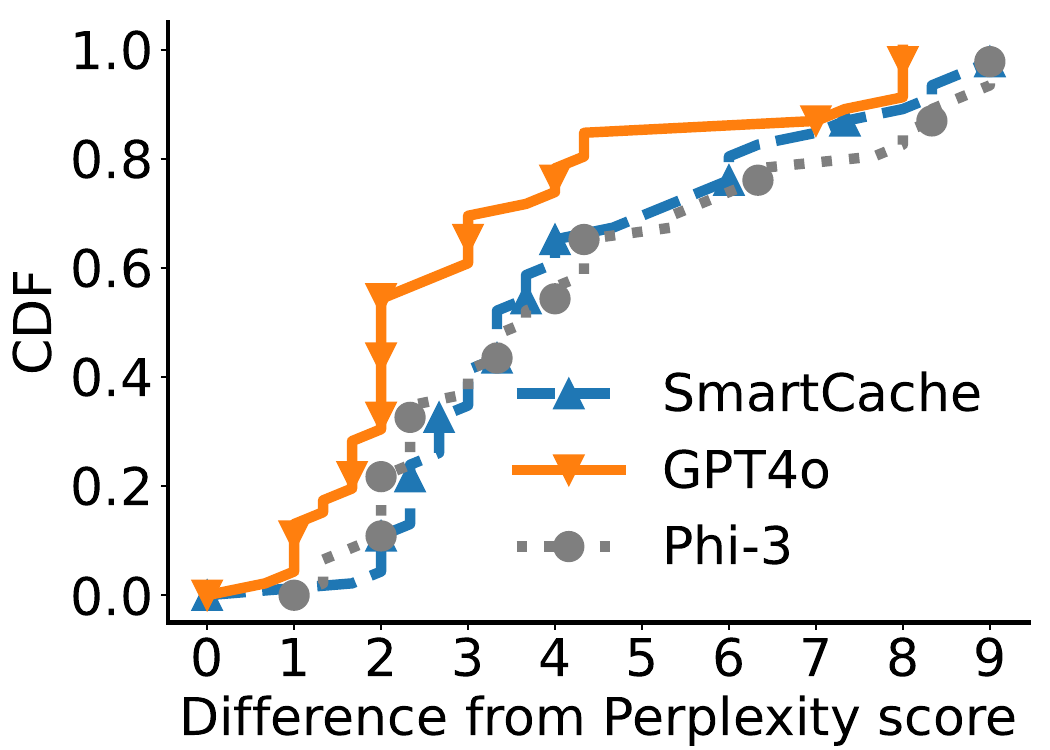}
    \label{subfig:cache_overall}}
    \subfloat[][Subset of queries]{\includegraphics[width=0.48\columnwidth]{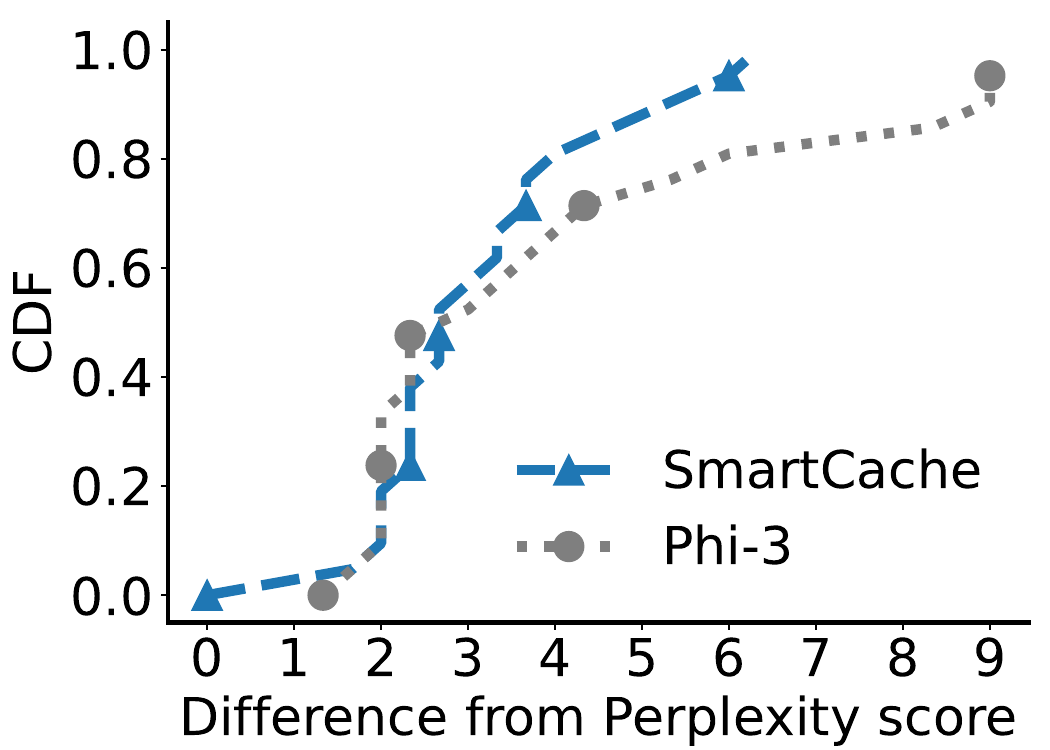}
    \label{subfig:cache_zoomed}}
    \caption{\ref{subfig:cache_overall} shows the quality CDF of the \texttt{smart\_cache} vs. directly using GPT4o/Phi-3. \ref{subfig:cache_zoomed} highlights the benefit of using factual information for smaller models which have a propensity to hallucinate.}
    \label{fig:cache_expt}
    % \vspace{-1.5em}
\end{figure}

\textit{Results.}
The results of our experiment are shown in Fig.~\ref{fig:cache_expt}.
We first focus on the overall quality of responses generated by different strategies (Fig.~\ref{subfig:cache_overall}).
GPT4o is considerably superior compared to using Phi-3 (as expected) for the majority of the factual queries, the worst-case being $\sim$8pts from the reference.
\texttt{smart\_cache} is able to bridge the quality gap, in particular for 20\% of the queries with the lowest quality.
While marginal, the improvement using \texttt{smart\_cache} is the difference between a factually sound (but a less detailed/creative) answer and a hallucinated (those generated by using Phi-3 alone) one.

To emphasize the benefits of leveraging the cached content we narrow down on the subset of queries where \texttt{smart\_cache} decides to use the cached information in Fig.~\ref{subfig:cache_zoomed}.
For such queries, \texttt{smart\_cache} has a considerable advantage over using Phi-3 in isolation --- the lowest score achieved on this subset by \texttt{smart\_cache} is 4pts vs. 1pt when using Phi-3 alone.
% One such query is: ``What is Dr. Miami's real name?''.
% \texttt{smart\_cache} utilizes the appropriate Wikipedia article in order to generate a response, without which Phi-3 hallucinates.
% The set of factual queries contain questions like:

% achieves the highest quality across all percentiles

% We compare the quality of responses generated by SmartCache with those generated by an expensive (GPT4o) and inexpensive LLM (Phi-3).

\section{Related Work}\label{sec:related-work}

% \fahad{Maybe the paragraphs could be: Abstractions (langchain, bedrock, openAI, parrot, etc), Model Selection, Caching, Context Management, Other Optimizations, Proxy}

\textbf{Abstractions:}
Systems such as Parrot~\cite{parrot_osdi} and Teola~\cite{teola} propose a more expressive LLM API that reveals dependencies between requests allowing for application level optimizations rather than request level.
Our proxy interfaces with existing LLM APIs and focuses specifically on cost optimizations that do not require modification in the LLM serving infrastructure.
Other abstractions such as LangChain~\cite{langchain} provide many building blocks, several of which can benefit \sys{} such as context summarizing (\S\ref{sec:design_context_manager}), to build LLM applications.
However, it does not provide a high level API like ours, requiring applications to figure out the appropriate configurations.
% \fahad{Maybe say that it doesn't provide model selection mechanisms nor a high level API like ours??}

\textbf{Model routing:} The problem of selecting the right LLM for a task is an active area of research, with many concurrent works, such as HybridLLM~\cite{ding2024hybrid}, RouteLLM~\cite{ong2024routellmlearningroutellms}, FrugalGPT~\cite{frugalgpt} and using benchmarks~\cite{shnitzer2024large} all involving a ``router'' to select the best model for a task. LLMBlender~\cite{llm-blender-2023}, which combines the strengths of multiple LLMs, and others have studied cascading approaches similar to our strategy in \S\ref{sec:design_model_adapter}~\cite{zhang2024llmcascademultiobjectiveoptimal, gupta2024languagemodelcascadestokenlevel}.
These strategies can potentially be made part of \sys{}'s design as well.
Our approach is simplistic and meant to highlight the design of \sys{} while not requiring any additional model training (which can be expensive) and still performing well on our production dataset.
Future work could provide a quantitative evaluation of the pros-and-cons of these different approaches and more insights into what kind of workloads a given strategy should be used for.

% The strategy we propose complements these by providing similar options via off-the-shelf model APIs and making them available through the proxy interface.
% Offloading to cloud APIs can be essential for some applications that have limited access to custom training or compute resources.

\textbf{Context management:} Other works lower cost by reducing the number of input tokens through models trained for this purpose~\cite{shekhar2024optimizingcostsllmusage, TCRA}. This could work in tandem with our SmartContext strategy as another context filter. While \sys{} targets QnA style LLM uses, other systems have more complex context management requirements such as generative agents~\cite{generative_agents}. They treat context as a ``memory stream'' that surfaces relevant memories for new queries. With some modification we believe our filter based API can also work for this style of context.

\textbf{Caching:} Systems such as GPTCache~\cite{bang-2023-gptcache} and MeanCache~\cite{meancache} use embedding models to reply to LLM queries with saved responses. Others have improved on the embedding models for more effective caching~\cite{zhu2024efficientpromptcachingembedding} and used LLMs to generate test inputs for semantic caches~\cite{llm_test_input}. Our interface for \sys{} is flexible enough to benefit from these efforts, and can also accommodate our strategy of intelligently populating the cache with high quality factual knowledge and using an inexpensive LLM to respond to user queries~(\S\ref{sec:design_cache}).
% populating the cache with high quality factual knowledge that can be used to respond to a user query.

\textbf{Other optimizations:} There have been other recent works that optimize aspects of LLM scheduling ~\cite{orca, fairness_osdi} and caching intermediate computation~\cite{ragcache} which can also benefit LLM APIs when they are used by \sys{}. Benchmarking model quality is also a recent area of research and we use LLM as a judge, inspired by~\cite{zheng2023judging}.

\textbf{Proxies:} There are many examples of performance optimizing proxies, some of which are primarily meant to reduce cost~\cite{catnap, polaris, flywheel, prophecy}. Others work at the transport level to improve performance~\cite{pep_satellite, rfc3135}, and others take into account application specific knowledge to improve performance~\cite{remote-control-caching, appex_conext}. These use optimizations such as caching and prefetching, which, with modification, can be used to improve LLM usage.

\section{Conclusion}

We introduced \sys{}, a proxy for supporting prompt-centric communication.
As a starting point, \sys{} focuses on cost-optimizations, offering a suitable interface to capture application preferences and internally using model selection, context management, and caching. Our design, implementation, and evaluation highlight the quantitative and qualitative benefits of our approach, in supporting rich services such as the WhatsApp Q\&A service, and diverse use-cases as well as providing cost benefits in various scenarios. 

\section{Acknowledgements}
Thanks to the Tufts NAT lab and D.O.C.C. lab as well as all the users of our WhatsApp Q\&A service for their feedback and support.

This work was partially funded by NSF CNS award: 2106797.

\bibliographystyle{ACM-Reference-Format}
\bibliography{main}


\begin{thebibliography}{74}

%%% the Web address.

\ifx \showCODEN    \undefined \def \showCODEN     #1{\unskip}     \fi
\ifx \showISBNx    \undefined \def \showISBNx     #1{\unskip}     \fi
\ifx \showISBNxiii \undefined \def \showISBNxiii  #1{\unskip}     \fi
\ifx \showISSN     \undefined \def \showISSN      #1{\unskip}     \fi
\ifx \showLCCN     \undefined \def \showLCCN      #1{\unskip}     \fi
\ifx \shownote     \undefined \def \shownote      #1{#1}          \fi
\ifx \showarticletitle \undefined \def \showarticletitle #1{#1}   \fi
\ifx \showURL      \undefined \def \showURL       {\relax}        \fi
% invisible to TeX
\providecommand\bibfield[2]{#2}
\providecommand\bibinfo[2]{#2}
\providecommand\natexlab[1]{#1}
\providecommand\showeprint[2][]{arXiv:#2}

\bibitem[alp(2023)]%
        {alphabet_reuters}
 \bibinfo{year}{{2023}}\natexlab{}.
\newblock \bibinfo{title}{{Focus: For tech giants, AI like Bing and Bard poses billion-dollar search problem}}.
\newblock \bibinfo{howpublished}{\url{https://www.reuters.com/technology/tech-giants-ai-like-bing-bard-poses-billion-dollar-search-problem-2023-02-22/}}.
\newblock


\bibitem[loc(2023)]%
        {local_llm}
 \bibinfo{year}{2023}\natexlab{}.
\newblock \bibinfo{title}{{No GPU? No problem. localllm lets you develop gen AI apps on local CPUs}}.
\newblock \bibinfo{howpublished}{\url{https://cloud.google.com/blog/products/application-development/new-localllm-lets-you-develop-gen-ai-apps-locally-without-gpus}}.
\newblock


\bibitem[ruf(2024)]%
        {rufusTechCrunch}
 \bibinfo{year}{2024}\natexlab{}.
\newblock \bibinfo{title}{{Amazon debuts ‘Rufus,’ an AI shopping assistant in its mobile app}}.
\newblock \bibinfo{howpublished}{\url{https://techcrunch.com/2024/02/01/amazon-debuts-rufus-an-ai-shopping-assistant-in-its-mobile-app/}}.
\newblock


\bibitem[ope(2024a)]%
        {openai_assistant}
 \bibinfo{year}{2024}\natexlab{a}.
\newblock \bibinfo{title}{{Assitants overview - OpenAI API}}.
\newblock \bibinfo{howpublished}{\url{https://platform.openai.com/docs/assistants/overview}}.
\newblock


\bibitem[bed(2024)]%
        {bedrock}
 \bibinfo{year}{2024}\natexlab{}.
\newblock \bibinfo{title}{{Build Generative AI Applications with Foundation Models - Amazon Bedrock - AWS}}.
\newblock \bibinfo{howpublished}{\url{https://aws.amazon.com/bedrock/}}.
\newblock


\bibitem[per(2024)]%
        {perplexityNYT}
 \bibinfo{year}{2024}\natexlab{}.
\newblock \bibinfo{title}{{Can This A.I.-Powered Search Engine Replace Google? It Has for Me.}}
\newblock \bibinfo{howpublished}{\url{https://www.nytimes.com/2024/02/01/technology/perplexity-search-ai-google.html}}.
\newblock


\bibitem[ope(2024b)]%
        {openai_pricing}
 \bibinfo{year}{2024}\natexlab{b}.
\newblock \bibinfo{title}{{ChatGPT Pricing | OpenAI}}.
\newblock \bibinfo{howpublished}{\url{https://openai.com/chatgpt/pricing/}}.
\newblock


\bibitem[aws(2024)]%
        {aws}
 \bibinfo{year}{2024}\natexlab{}.
\newblock \bibinfo{title}{{Cloud Computing Services - Amazon Web Services (AWS)}}.
\newblock \bibinfo{howpublished}{\url{https://aws.amazon.com}}.
\newblock


\bibitem[ope(2024c)]%
        {openai_embeddings}
 \bibinfo{year}{2024}\natexlab{c}.
\newblock \bibinfo{title}{{Embeddings - OpenAI API}}.
\newblock \bibinfo{howpublished}{\url{https://platform.openai.com/docs/guides/embeddings}}.
\newblock


\bibitem[phi(2024)]%
        {phi3}
 \bibinfo{year}{2024}\natexlab{}.
\newblock \bibinfo{title}{{Introducing Phi-3: Redefining what’s possible with SLMs }}.
\newblock \bibinfo{howpublished}{\url{https://azure.microsoft.com/en-us/blog/introducing-phi-3-redefining-whats-possible-with-slms/}}.
\newblock


\bibitem[ope(2024d)]%
        {openai_docs}
 \bibinfo{year}{2024}\natexlab{d}.
\newblock \bibinfo{title}{{Introduction - OpenAI API}}.
\newblock \bibinfo{howpublished}{\url{https://platform.openai.com/docs/introduction}}.
\newblock


\bibitem[llm(2024)]%
        {llm_economics}
 \bibinfo{year}{2024}\natexlab{}.
\newblock \bibinfo{title}{{LLM Economics: ChatGPT vs Open-Source}}.
\newblock \bibinfo{howpublished}{\url{https://towardsdatascience.com/llm-economics-chatgpt-vs-open-source-dfc29f69fec1}}.
\newblock


\bibitem[ant(2024)]%
        {anthropic}
 \bibinfo{year}{2024}\natexlab{}.
\newblock \bibinfo{title}{{Meet Claude \ Anthropic}}.
\newblock \bibinfo{howpublished}{\url{https://www.anthropic.com/claude}}.
\newblock


\bibitem[azu(2024)]%
        {azure_openai_pricing}
 \bibinfo{year}{2024}\natexlab{}.
\newblock \bibinfo{title}{OpenAI Service Pricing - Microsoft Azure}.
\newblock \bibinfo{howpublished}{\url{https://azure.microsoft.com/en-us/pricing/details/cognitive-services/openai-service/}}.
\newblock
\newblock
\shownote{Accessed: 2024-04-30}.


\bibitem[wha(2024a)]%
        {whatsapp}
 \bibinfo{year}{2024}\natexlab{a}.
\newblock \bibinfo{title}{{WhatsApp | Secure and Reliable Free Private Messaging and Calling}}.
\newblock \bibinfo{howpublished}{\url{https://www.whatsapp.com}}.
\newblock


\bibitem[wha(2024b)]%
        {whatsapp-popularity}
 \bibinfo{year}{2024}\natexlab{b}.
\newblock \bibinfo{title}{{WhatsApp Users by Country 2024}}.
\newblock \bibinfo{howpublished}{\url{https://worldpopulationreview.com/country-rankings/whatsapp-users-by-country}}.
\newblock


\bibitem[wik(2024)]%
        {wikipedia}
 \bibinfo{year}{2024}\natexlab{}.
\newblock \bibinfo{title}{{Wikipedia}}.
\newblock \bibinfo{howpublished}{\url{https://www.wikipedia.org/}}.
\newblock


\bibitem[azu(2025)]%
        {azure_availability}
 \bibinfo{year}{2025}\natexlab{}.
\newblock \bibinfo{title}{Azure OpenAI Service models}.
\newblock \bibinfo{howpublished}{\url{https://learn.microsoft.com/en-us/azure/ai-services/openai/concepts/models}}.
\newblock


\bibitem[lla(2025)]%
        {llama}
 \bibinfo{year}{2025}\natexlab{}.
\newblock \bibinfo{title}{Llama}.
\newblock \bibinfo{howpublished}{\url{https://www.llama.com/}}.
\newblock


\bibitem[bed(2025a)]%
        {bedrock_availability}
 \bibinfo{year}{2025}\natexlab{a}.
\newblock \bibinfo{title}{Model support by AWS Region in Amazon Bedrock}.
\newblock \bibinfo{howpublished}{\url{https://docs.aws.amazon.com/bedrock/latest/userguide/models-regions.html}}.
\newblock


\bibitem[per(2025)]%
        {perplexity}
 \bibinfo{year}{2025}\natexlab{}.
\newblock \bibinfo{title}{{Perplexity Sonar Models}}.
\newblock \bibinfo{howpublished}{\url{https://docs.perplexity.ai/guides/model-cards\#perplexity-sonar-models/}}.
\newblock


\bibitem[phi(2025)]%
        {phi}
 \bibinfo{year}{2025}\natexlab{}.
\newblock \bibinfo{title}{Phi}.
\newblock \bibinfo{howpublished}{\url{https://azure.microsoft.com/en-us/products/phi}}.
\newblock


\bibitem[bed(2025b)]%
        {bedrock-routing}
 \bibinfo{year}{2025}\natexlab{b}.
\newblock \bibinfo{title}{Route Prompts Between Models - Amazon Bedrock Intelligent Prompt Routing - AWS}.
\newblock \bibinfo{howpublished}{\url{https://aws.amazon.com/bedrock/intelligent-prompt-routing/}}.
\newblock


\bibitem[Agababov et~al\mbox{.}(2015)]%
        {flywheel}
\bibfield{author}{\bibinfo{person}{Victor Agababov}, \bibinfo{person}{Michael Buettner}, \bibinfo{person}{Victor Chudnovsky}, \bibinfo{person}{Mark Cogan}, \bibinfo{person}{Ben Greenstein}, \bibinfo{person}{Shane McDaniel}, \bibinfo{person}{Michael Piatek}, \bibinfo{person}{Colin Scott}, \bibinfo{person}{Matt Welsh}, {and} \bibinfo{person}{Bolian Yin}.} \bibinfo{year}{2015}\natexlab{}.
\newblock \showarticletitle{Flywheel: {Google{\textquoteright}s} Data Compression Proxy for the Mobile Web}. In \bibinfo{booktitle}{\emph{12th USENIX Symposium on Networked Systems Design and Implementation (NSDI 15)}}. \bibinfo{publisher}{USENIX Association}, \bibinfo{address}{Oakland, CA}, \bibinfo{pages}{367--380}.
\newblock
\showISBNx{978-1-931971-218}
\urldef\tempurl%
\url{https://www.usenix.org/conference/nsdi15/technical-sessions/presentation/agababov}
\showURL{%
\tempurl}


\bibitem[Arawjo et~al\mbox{.}(2024)]%
        {chainforge}
\bibfield{author}{\bibinfo{person}{Ian Arawjo}, \bibinfo{person}{Chelse Swoopes}, \bibinfo{person}{Priyan Vaithilingam}, \bibinfo{person}{Martin Wattenberg}, {and} \bibinfo{person}{Elena~L. Glassman}.} \bibinfo{year}{2024}\natexlab{}.
\newblock \showarticletitle{ChainForge: A Visual Toolkit for Prompt Engineering and LLM Hypothesis Testing}. In \bibinfo{booktitle}{\emph{Proceedings of the 2024 CHI Conference on Human Factors in Computing Systems}} \emph{(\bibinfo{series}{CHI '24})}. \bibinfo{publisher}{Association for Computing Machinery}.
\newblock


\bibitem[Bang(2023)]%
        {bang-2023-gptcache}
\bibfield{author}{\bibinfo{person}{Fu Bang}.} \bibinfo{year}{2023}\natexlab{}.
\newblock \showarticletitle{{GPTC}ache: An Open-Source Semantic Cache for {LLM} Applications Enabling Faster Answers and Cost Savings}. In \bibinfo{booktitle}{\emph{Proceedings of the 3rd Workshop for Natural Language Processing Open Source Software (NLP-OSS 2023)}}, \bibfield{editor}{\bibinfo{person}{Liling Tan}, \bibinfo{person}{Dmitrijs Milajevs}, \bibinfo{person}{Geeticka Chauhan}, \bibinfo{person}{Jeremy Gwinnup}, {and} \bibinfo{person}{Elijah Rippeth}} (Eds.). \bibinfo{publisher}{Association for Computational Linguistics}, \bibinfo{address}{Singapore}.
\newblock
\href{https://doi.org/10.18653/v1/2023.nlposs-1.24}{doi:\nolinkurl{10.18653/v1/2023.nlposs-1.24}}


\bibitem[Caini et~al\mbox{.}(2006)]%
        {pep_satellite}
\bibfield{author}{\bibinfo{person}{C. Caini}, \bibinfo{person}{R. Firrincieli}, {and} \bibinfo{person}{D. Lacamera}.} \bibinfo{year}{2006}\natexlab{}.
\newblock \showarticletitle{PEPsal: a Performance Enhancing Proxy designed for TCP satellite connections}. In \bibinfo{booktitle}{\emph{2006 IEEE 63rd Vehicular Technology Conference}}, Vol.~\bibinfo{volume}{6}. \bibinfo{pages}{2607--2611}.
\newblock
\href{https://doi.org/10.1109/VETECS.2006.1683339}{doi:\nolinkurl{10.1109/VETECS.2006.1683339}}


\bibitem[Chase(2022)]%
        {langchain}
\bibfield{author}{\bibinfo{person}{Harrison Chase}.} \bibinfo{year}{2022}\natexlab{}.
\newblock \bibinfo{title}{{LangChain}}.
\newblock
\urldef\tempurl%
\url{https://github.com/langchain-ai/langchain}
\showURL{%
\tempurl}


\bibitem[Chen et~al\mbox{.}(2023)]%
        {frugalgpt}
\bibfield{author}{\bibinfo{person}{Lingjiao Chen}, \bibinfo{person}{Matei Zaharia}, {and} \bibinfo{person}{James Zou}.} \bibinfo{year}{2023}\natexlab{}.
\newblock \showarticletitle{FrugalGPT: How to Use Large Language Models While Reducing Cost and Improving Performance}.
\newblock \bibinfo{journal}{\emph{arXiv preprint arXiv:2305.05176}} (\bibinfo{year}{2023}).
\newblock
\showeprint{2305.05176}~[cs.LG]
\urldef\tempurl%
\url{https://arxiv.org/abs/2305.05176}
\showURL{%
\tempurl}


\bibitem[Choi et~al\mbox{.}(2018)]%
        {appex_conext}
\bibfield{author}{\bibinfo{person}{Byungkwon Choi}, \bibinfo{person}{Jeongmin Kim}, \bibinfo{person}{Daeyang Cho}, \bibinfo{person}{Seongmin Kim}, {and} \bibinfo{person}{Dongsu Han}.} \bibinfo{year}{2018}\natexlab{}.
\newblock \showarticletitle{APPx: an automated app acceleration framework for low latency mobile app}. In \bibinfo{booktitle}{\emph{Proceedings of the 14th International Conference on Emerging Networking EXperiments and Technologies}} (Heraklion, Greece) \emph{(\bibinfo{series}{CoNEXT '18})}. \bibinfo{publisher}{Association for Computing Machinery}, \bibinfo{address}{New York, NY, USA}, \bibinfo{pages}{27–40}.
\newblock
\showISBNx{9781450360807}
\href{https://doi.org/10.1145/3281411.3281416}{doi:\nolinkurl{10.1145/3281411.3281416}}


\bibitem[Ding et~al\mbox{.}(2024)]%
        {ding2024hybrid}
\bibfield{author}{\bibinfo{person}{Dujian Ding}, \bibinfo{person}{Ankur Mallick}, \bibinfo{person}{Chi Wang}, \bibinfo{person}{Robert Sim}, \bibinfo{person}{Subhabrata Mukherjee}, \bibinfo{person}{Victor Ruehle}, \bibinfo{person}{Laks V.~S. Lakshmanan}, {and} \bibinfo{person}{Ahmed Awadallah}.} \bibinfo{year}{2024}\natexlab{}.
\newblock \showarticletitle{Hybrid LLM: Cost-Efficient and Quality-Aware Query Routing}. In \bibinfo{booktitle}{\emph{ICLR 2024}}.
\newblock
\urldef\tempurl%
\url{https://www.microsoft.com/en-us/research/publication/hybrid-llm-cost-efficient-and-quality-aware-query-routing/}
\showURL{%
\tempurl}


\bibitem[Dogar et~al\mbox{.}(2020)]%
        {missit}
\bibfield{author}{\bibinfo{person}{Fahad~R Dogar}, \bibinfo{person}{Ihsan~Ayyub Qazi}, \bibinfo{person}{Ali~Raza Tariq}, \bibinfo{person}{Ghulam Murtaza}, \bibinfo{person}{Abeer Ahmad}, {and} \bibinfo{person}{Nathan Stocking}.} \bibinfo{year}{2020}\natexlab{}.
\newblock \showarticletitle{Missit: Using missed calls for free, extremely low bit-rate communication in developing regions}. In \bibinfo{booktitle}{\emph{Proceedings of the 2020 CHI Conference on Human Factors in Computing Systems (CHI)}}. \bibinfo{pages}{1--12}.
\newblock


\bibitem[Dogar and Steenkiste(2012)]%
        {tapa}
\bibfield{author}{\bibinfo{person}{Fahad~R. Dogar} {and} \bibinfo{person}{Peter Steenkiste}.} \bibinfo{year}{2012}\natexlab{}.
\newblock \showarticletitle{Architecting for edge diversity: supporting rich services over an unbundled transport}. In \bibinfo{booktitle}{\emph{Proceedings of the 8th International Conference on Emerging Networking Experiments and Technologies}} (Nice, France) \emph{(\bibinfo{series}{CoNEXT '12})}. \bibinfo{publisher}{Association for Computing Machinery}, \bibinfo{address}{New York, NY, USA}, \bibinfo{pages}{13–24}.
\newblock
\showISBNx{9781450317757}
\href{https://doi.org/10.1145/2413176.2413179}{doi:\nolinkurl{10.1145/2413176.2413179}}


\bibitem[Dogar et~al\mbox{.}(2010)]%
        {catnap}
\bibfield{author}{\bibinfo{person}{Fahad~R. Dogar}, \bibinfo{person}{Peter Steenkiste}, {and} \bibinfo{person}{Konstantina Papagiannaki}.} \bibinfo{year}{2010}\natexlab{}.
\newblock \showarticletitle{Catnap: exploiting high bandwidth wireless interfaces to save energy for mobile devices}. In \bibinfo{booktitle}{\emph{Proceedings of the 8th International Conference on Mobile Systems, Applications, and Services}} (San Francisco, California, USA) \emph{(\bibinfo{series}{MobiSys '10})}. \bibinfo{publisher}{Association for Computing Machinery}, \bibinfo{address}{New York, NY, USA}, \bibinfo{pages}{107–122}.
\newblock
\showISBNx{9781605589855}
\href{https://doi.org/10.1145/1814433.1814446}{doi:\nolinkurl{10.1145/1814433.1814446}}


\bibitem[Durumeric et~al\mbox{.}(2017)]%
        {durumeric2017security}
\bibfield{author}{\bibinfo{person}{Zakir Durumeric}, \bibinfo{person}{Zane Ma}, \bibinfo{person}{Drew Springall}, \bibinfo{person}{Richard Barnes}, \bibinfo{person}{Nick Sullivan}, \bibinfo{person}{Elie Bursztein}, \bibinfo{person}{Michael~D Bailey}, \bibinfo{person}{J~Alex Halderman}, {and} \bibinfo{person}{Vern Paxson}.} \bibinfo{year}{2017}\natexlab{}.
\newblock \showarticletitle{The Security Impact of HTTPS Interception.}. In \bibinfo{booktitle}{\emph{NDSS}}.
\newblock


\bibitem[Gandhi et~al\mbox{.}(2016)]%
        {yodaLoadBalancer}
\bibfield{author}{\bibinfo{person}{Rohan Gandhi}, \bibinfo{person}{Y.~Charlie Hu}, {and} \bibinfo{person}{Ming Zhang}.} \bibinfo{year}{2016}\natexlab{}.
\newblock \showarticletitle{Yoda: a highly available layer-7 load balancer}. In \bibinfo{booktitle}{\emph{Proceedings of the Eleventh European Conference on Computer Systems}} (London, United Kingdom) \emph{(\bibinfo{series}{EuroSys '16})}. \bibinfo{publisher}{Association for Computing Machinery}, \bibinfo{address}{New York, NY, USA}, Article \bibinfo{articleno}{21}, \bibinfo{numpages}{16}~pages.
\newblock
\showISBNx{9781450342407}
\href{https://doi.org/10.1145/2901318.2901352}{doi:\nolinkurl{10.1145/2901318.2901352}}


\bibitem[Gao et~al\mbox{.}(2022)]%
        {hyde}
\bibfield{author}{\bibinfo{person}{Luyu Gao}, \bibinfo{person}{Xueguang Ma}, \bibinfo{person}{Jimmy Lin}, {and} \bibinfo{person}{Jamie Callan}.} \bibinfo{year}{2022}\natexlab{}.
\newblock \showarticletitle{Precise Zero-Shot Dense Retrieval without Relevance Labels}.
\newblock \bibinfo{journal}{\emph{arXiv preprint arXiv:2212.10496}} (\bibinfo{year}{2022}).
\newblock


\bibitem[Gill et~al\mbox{.}(2024)]%
        {meancache}
\bibfield{author}{\bibinfo{person}{Waris Gill}, \bibinfo{person}{Mohamed Elidrisi}, \bibinfo{person}{Pallavi Kalapatapu}, \bibinfo{person}{Ali Anwar}, {and} \bibinfo{person}{Muhammad~Ali Gulzar}.} \bibinfo{year}{2024}\natexlab{}.
\newblock \showarticletitle{Privacy-Aware Semantic Cache for Large Language Models}.
\newblock \bibinfo{journal}{\emph{arXiv preprint arXiv:2403.02694}} (\bibinfo{year}{2024}).
\newblock


\bibitem[{Google Cloud}(2025)]%
        {gemini_2_0_flash}
\bibfield{author}{\bibinfo{person}{{Google Cloud}}.} \bibinfo{year}{2025}\natexlab{}.
\newblock \bibinfo{title}{Gemini 2.0 Flash}.
\newblock \bibinfo{howpublished}{\url{https://cloud.google.com/vertex-ai/generative-ai/docs/models/gemini/2-0-flash}}.
\newblock


\bibitem[Griner et~al\mbox{.}(2001)]%
        {rfc3135}
\bibfield{author}{\bibinfo{person}{Jim Griner}, \bibinfo{person}{John Border}, \bibinfo{person}{Markku Kojo}, \bibinfo{person}{Zach~D. Shelby}, {and} \bibinfo{person}{Gabriel Montenegro}.} \bibinfo{year}{2001}\natexlab{}.
\newblock \bibinfo{title}{{Performance Enhancing Proxies Intended to Mitigate Link-Related Degradations}}.
\newblock \bibinfo{howpublished}{RFC 3135}.
\newblock
\href{https://doi.org/10.17487/RFC3135}{doi:\nolinkurl{10.17487/RFC3135}}


\bibitem[Gupta et~al\mbox{.}(2024)]%
        {gupta2024languagemodelcascadestokenlevel}
\bibfield{author}{\bibinfo{person}{Neha Gupta}, \bibinfo{person}{Harikrishna Narasimhan}, \bibinfo{person}{Wittawat Jitkrittum}, \bibinfo{person}{Ankit~Singh Rawat}, \bibinfo{person}{Aditya~Krishna Menon}, {and} \bibinfo{person}{Sanjiv Kumar}.} \bibinfo{year}{2024}\natexlab{}.
\newblock \bibinfo{title}{Language Model Cascades: Token-level uncertainty and beyond}.
\newblock
\showeprint[arxiv]{2404.10136}~[cs.CL]
\urldef\tempurl%
\url{https://arxiv.org/abs/2404.10136}
\showURL{%
\tempurl}


\bibitem[Haq et~al\mbox{.}(2017)]%
        {cloud_paths_www}
\bibfield{author}{\bibinfo{person}{Osama Haq}, \bibinfo{person}{Mamoon Raja}, {and} \bibinfo{person}{Fahad~R. Dogar}.} \bibinfo{year}{2017}\natexlab{}.
\newblock \showarticletitle{Measuring and Improving the Reliability of Wide-Area Cloud Paths}. In \bibinfo{booktitle}{\emph{Proceedings of the 26th International Conference on World Wide Web}} (Perth, Australia) \emph{(\bibinfo{series}{WWW '17})}. \bibinfo{publisher}{International World Wide Web Conferences Steering Committee}, \bibinfo{address}{Republic and Canton of Geneva, CHE}, \bibinfo{pages}{253–262}.
\newblock
\showISBNx{9781450349130}
\href{https://doi.org/10.1145/3038912.3052560}{doi:\nolinkurl{10.1145/3038912.3052560}}


\bibitem[Haroon and Dogar(2024)]%
        {twips-haroon}
\bibfield{author}{\bibinfo{person}{Rukhshan Haroon} {and} \bibinfo{person}{Fahad Dogar}.} \bibinfo{year}{2024}\natexlab{}.
\newblock \showarticletitle{TwIPS: A Large Language Model Powered Texting Application to Simplify Conversational Nuances for Autistic Users}. In \bibinfo{booktitle}{\emph{Proceedings of the 26th International ACM SIGACCESS Conference on Computers and Accessibility (to appear)}}.
\newblock
\urldef\tempurl%
\url{https://rukhshan23.github.io/twips.pdf}
\showURL{%
\tempurl}


\bibitem[He et~al\mbox{.}(2021)]%
        {he2021automl}
\bibfield{author}{\bibinfo{person}{Xin He}, \bibinfo{person}{Kaiyong Zhao}, {and} \bibinfo{person}{Xiaowen Chu}.} \bibinfo{year}{2021}\natexlab{}.
\newblock \showarticletitle{AutoML: A survey of the state-of-the-art}.
\newblock \bibinfo{journal}{\emph{Knowledge-based systems}}  \bibinfo{volume}{212} (\bibinfo{year}{2021}), \bibinfo{pages}{106622}.
\newblock


\bibitem[He et~al\mbox{.}(2024)]%
        {llmTranslation}
\bibfield{author}{\bibinfo{person}{Zhiwei He}, \bibinfo{person}{Tian Liang}, \bibinfo{person}{Wenxiang Jiao}, \bibinfo{person}{Zhuosheng Zhang}, \bibinfo{person}{Yujiu Yang}, \bibinfo{person}{Rui Wang}, \bibinfo{person}{Zhaopeng Tu}, \bibinfo{person}{Shuming Shi}, {and} \bibinfo{person}{Xing Wang}.} \bibinfo{year}{2024}\natexlab{}.
\newblock \showarticletitle{Exploring Human-Like Translation Strategy with Large Language Models}.
\newblock \bibinfo{journal}{\emph{Transactions of the Association for Computational Linguistics}}  \bibinfo{volume}{12} (\bibinfo{date}{03} \bibinfo{year}{2024}), \bibinfo{pages}{229--246}.
\newblock
\showISSN{2307-387X}
\href{https://doi.org/10.1162/tacl_a_00642}{doi:\nolinkurl{10.1162/tacl_a_00642}}
\showeprint{https://direct.mit.edu/tacl/article-pdf/doi/10.1162/tacl\_a\_00642/2346100/tacl\_a\_00642.pdf}


\bibitem[Jahanbakhsh et~al\mbox{.}(2023)]%
        {ai-for-misinformation}
\bibfield{author}{\bibinfo{person}{Farnaz Jahanbakhsh}, \bibinfo{person}{Yannis Katsis}, \bibinfo{person}{Dakuo Wang}, \bibinfo{person}{Lucian Popa}, {and} \bibinfo{person}{Michael Muller}.} \bibinfo{year}{2023}\natexlab{}.
\newblock \showarticletitle{Exploring the Use of Personalized AI for Identifying Misinformation on Social Media} \emph{(\bibinfo{series}{CHI '23})}. \bibinfo{publisher}{Association for Computing Machinery}.
\newblock


\bibitem[Jiang et~al\mbox{.}(2023)]%
        {llm-blender-2023}
\bibfield{author}{\bibinfo{person}{Dongfu Jiang}, \bibinfo{person}{Xiang Ren}, {and} \bibinfo{person}{Bill~Yuchen Lin}.} \bibinfo{year}{2023}\natexlab{}.
\newblock \showarticletitle{LLM-Blender: Ensembling Large Language Models with Pairwise Comparison and Generative Fusion}. In \bibinfo{booktitle}{\emph{Proceedings of the 61th Annual Meeting of the Association for Computational Linguistics (ACL 2023)}}.
\newblock


\bibitem[Jin et~al\mbox{.}(2024)]%
        {ragcache}
\bibfield{author}{\bibinfo{person}{Chao Jin}, \bibinfo{person}{Zili Zhang}, \bibinfo{person}{Xuanlin Jiang}, \bibinfo{person}{Fangyue Liu}, \bibinfo{person}{Xin Liu}, \bibinfo{person}{Xuanzhe Liu}, {and} \bibinfo{person}{Xin Jin}.} \bibinfo{year}{2024}\natexlab{}.
\newblock \showarticletitle{RAGCache: Efficient Knowledge Caching for Retrieval-Augmented Generation}.
\newblock \bibinfo{journal}{\emph{arXiv preprint arXiv:2404.12457}} (\bibinfo{year}{2024}).
\newblock
\showeprint{2404.12457}~[cs.DC]
\urldef\tempurl%
\url{https://arxiv.org/abs/2404.12457}
\showURL{%
\tempurl}


\bibitem[Lewis et~al\mbox{.}(2021)]%
        {RAG}
\bibfield{author}{\bibinfo{person}{Patrick Lewis}, \bibinfo{person}{Ethan Perez}, \bibinfo{person}{Aleksandra Piktus}, \bibinfo{person}{Fabio Petroni}, \bibinfo{person}{Vladimir Karpukhin}, \bibinfo{person}{Naman Goyal}, \bibinfo{person}{Heinrich Küttler}, \bibinfo{person}{Mike Lewis}, \bibinfo{person}{Wen tau Yih}, \bibinfo{person}{Tim Rocktäschel}, \bibinfo{person}{Sebastian Riedel}, {and} \bibinfo{person}{Douwe Kiela}.} \bibinfo{year}{2021}\natexlab{}.
\newblock \showarticletitle{Retrieval-Augmented Generation for Knowledge-Intensive NLP Tasks}.
\newblock \bibinfo{journal}{\emph{arXiv preprint arXiv:2005.11401}} (\bibinfo{year}{2021}).
\newblock
\showeprint{2005.11401}~[cs.CL]
\urldef\tempurl%
\url{https://arxiv.org/abs/2005.11401}
\showURL{%
\tempurl}


\bibitem[Lin et~al\mbox{.}(2024)]%
        {parrot_osdi}
\bibfield{author}{\bibinfo{person}{Chaofan Lin}, \bibinfo{person}{Zhenhua Han}, \bibinfo{person}{Chengruidong Zhang}, \bibinfo{person}{Yuqing Yang}, \bibinfo{person}{Fan Yang}, \bibinfo{person}{Chen Chen}, {and} \bibinfo{person}{Lili Qiu}.} \bibinfo{year}{2024}\natexlab{}.
\newblock \showarticletitle{Parrot: Efficient Serving of {LLM-based} Applications with Semantic Variable}. In \bibinfo{booktitle}{\emph{18th USENIX Symposium on Operating Systems Design and Implementation (OSDI 24)}}. \bibinfo{publisher}{USENIX Association}, \bibinfo{address}{Santa Clara, CA}, \bibinfo{pages}{929--945}.
\newblock
\showISBNx{978-1-939133-40-3}
\urldef\tempurl%
\url{https://www.usenix.org/conference/osdi24/presentation/lin-chaofan}
\showURL{%
\tempurl}


\bibitem[Lin et~al\mbox{.}(2025)]%
        {sleep-time-compute}
\bibfield{author}{\bibinfo{person}{Kevin Lin}, \bibinfo{person}{Charlie Snell}, \bibinfo{person}{Yu Wang}, \bibinfo{person}{Charles Packer}, \bibinfo{person}{Sarah Wooders}, \bibinfo{person}{Ion Stoica}, {and} \bibinfo{person}{Joseph~E Gonzalez}.} \bibinfo{year}{2025}\natexlab{}.
\newblock \showarticletitle{Sleep-time compute: Beyond inference scaling at test-time}.
\newblock \bibinfo{journal}{\emph{arXiv preprint arXiv:2504.13171}} (\bibinfo{year}{2025}).
\newblock


\bibitem[Liu et~al\mbox{.}(2023)]%
        {TCRA}
\bibfield{author}{\bibinfo{person}{Junyi Liu}, \bibinfo{person}{Liangzhi Li}, \bibinfo{person}{Tong Xiang}, \bibinfo{person}{Bowen Wang}, {and} \bibinfo{person}{Yiming Qian}.} \bibinfo{year}{2023}\natexlab{}.
\newblock \showarticletitle{{TCRA}-{LLM}: Token Compression Retrieval Augmented Large Language Model for Inference Cost Reduction}. In \bibinfo{booktitle}{\emph{Findings of the Association for Computational Linguistics: EMNLP 2023}}, \bibfield{editor}{\bibinfo{person}{Houda Bouamor}, \bibinfo{person}{Juan Pino}, {and} \bibinfo{person}{Kalika Bali}} (Eds.). \bibinfo{publisher}{Association for Computational Linguistics}, \bibinfo{address}{Singapore}, \bibinfo{pages}{9796--9810}.
\newblock
\href{https://doi.org/10.18653/v1/2023.findings-emnlp.655}{doi:\nolinkurl{10.18653/v1/2023.findings-emnlp.655}}


\bibitem[Martin and Dogar(2023)]%
        {divide-conext}
\bibfield{author}{\bibinfo{person}{Noah Martin} {and} \bibinfo{person}{Fahad Dogar}.} \bibinfo{year}{2023}\natexlab{}.
\newblock \showarticletitle{Divided at the Edge - Measuring Performance and the Digital Divide of Cloud Edge Data Centers}.
\newblock \bibinfo{journal}{\emph{Proc. ACM Netw.}} \bibinfo{volume}{1}, \bibinfo{number}{CoNEXT3}, Article \bibinfo{articleno}{16} (\bibinfo{date}{nov} \bibinfo{year}{2023}), \bibinfo{numpages}{23}~pages.
\newblock
\href{https://doi.org/10.1145/3629138}{doi:\nolinkurl{10.1145/3629138}}


\bibitem[Netravali et~al\mbox{.}(2016)]%
        {polaris}
\bibfield{author}{\bibinfo{person}{Ravi Netravali}, \bibinfo{person}{Ameesh Goyal}, \bibinfo{person}{James Mickens}, {and} \bibinfo{person}{Hari Balakrishnan}.} \bibinfo{year}{2016}\natexlab{}.
\newblock \showarticletitle{Polaris: Faster Page Loads Using Fine-grained Dependency Tracking}. In \bibinfo{booktitle}{\emph{13th USENIX Symposium on Networked Systems Design and Implementation (NSDI 16)}}. \bibinfo{publisher}{USENIX Association}, \bibinfo{address}{Santa Clara, CA}.
\newblock
\urldef\tempurl%
\url{https://www.usenix.org/conference/nsdi16/technical-sessions/presentation/netravali}
\showURL{%
\tempurl}


\bibitem[Netravali and Mickens(2018a)]%
        {prophecy}
\bibfield{author}{\bibinfo{person}{Ravi Netravali} {and} \bibinfo{person}{James Mickens}.} \bibinfo{year}{2018}\natexlab{a}.
\newblock \showarticletitle{Prophecy: accelerating mobile page loads using final-state write logs}. In \bibinfo{booktitle}{\emph{Proceedings of the 15th USENIX Conference on Networked Systems Design and Implementation}} (Renton, WA, USA) \emph{(\bibinfo{series}{NSDI'18})}. \bibinfo{publisher}{USENIX Association}, \bibinfo{address}{USA}, \bibinfo{pages}{249–266}.
\newblock
\showISBNx{9781931971430}


\bibitem[Netravali and Mickens(2018b)]%
        {remote-control-caching}
\bibfield{author}{\bibinfo{person}{Ravi Netravali} {and} \bibinfo{person}{James Mickens}.} \bibinfo{year}{2018}\natexlab{b}.
\newblock \showarticletitle{Remote-Control Caching: Proxy-based URL Rewriting to Decrease Mobile Browsing Bandwidth}. In \bibinfo{booktitle}{\emph{Proceedings of the 19th International Workshop on Mobile Computing Systems \& Applications}} (Tempe, Arizona, USA) \emph{(\bibinfo{series}{HotMobile '18})}. \bibinfo{publisher}{Association for Computing Machinery}, \bibinfo{address}{New York, NY, USA}, \bibinfo{pages}{63–68}.
\newblock
\showISBNx{9781450356305}
\href{https://doi.org/10.1145/3177102.3177118}{doi:\nolinkurl{10.1145/3177102.3177118}}


\bibitem[Oelen and Auer(2024)]%
        {llm-in-ui}
\bibfield{author}{\bibinfo{person}{Allard Oelen} {and} \bibinfo{person}{S\"{o}ren Auer}.} \bibinfo{year}{2024}\natexlab{}.
\newblock \showarticletitle{Leveraging Large Language Models for Realizing Truly Intelligent User Interfaces} \emph{(\bibinfo{series}{CHI EA '24})}. \bibinfo{publisher}{Association for Computing Machinery}.
\newblock


\bibitem[Ong et~al\mbox{.}(2024)]%
        {ong2024routellmlearningroutellms}
\bibfield{author}{\bibinfo{person}{Isaac Ong}, \bibinfo{person}{Amjad Almahairi}, \bibinfo{person}{Vincent Wu}, \bibinfo{person}{Wei-Lin Chiang}, \bibinfo{person}{Tianhao Wu}, \bibinfo{person}{Joseph~E. Gonzalez}, \bibinfo{person}{M~Waleed Kadous}, {and} \bibinfo{person}{Ion Stoica}.} \bibinfo{year}{2024}\natexlab{}.
\newblock \showarticletitle{RouteLLM: Learning to Route LLMs with Preference Data}.
\newblock \bibinfo{journal}{\emph{arXiv preprint arXiv:2406.18665}} (\bibinfo{year}{2024}).
\newblock
\showeprint{2406.18665}~[cs.LG]
\urldef\tempurl%
\url{https://arxiv.org/abs/2406.18665}
\showURL{%
\tempurl}


\bibitem[Park et~al\mbox{.}(2023)]%
        {generative_agents}
\bibfield{author}{\bibinfo{person}{Joon~Sung Park}, \bibinfo{person}{Joseph O'Brien}, \bibinfo{person}{Carrie~Jun Cai}, \bibinfo{person}{Meredith~Ringel Morris}, \bibinfo{person}{Percy Liang}, {and} \bibinfo{person}{Michael~S. Bernstein}.} \bibinfo{year}{2023}\natexlab{}.
\newblock \showarticletitle{Generative Agents: Interactive Simulacra of Human Behavior}. In \bibinfo{booktitle}{\emph{Proceedings of the 36th Annual ACM Symposium on User Interface Software and Technology}} (San Francisco, CA, USA) \emph{(\bibinfo{series}{UIST '23})}. \bibinfo{publisher}{Association for Computing Machinery}, \bibinfo{address}{New York, NY, USA}, Article \bibinfo{articleno}{2}, \bibinfo{numpages}{22}~pages.
\newblock
\showISBNx{9798400701320}
\urldef\tempurl%
\url{https://doi.org/10.1145/3586183.3606763}
\showURL{%
\tempurl}


\bibitem[Popa et~al\mbox{.}(2010)]%
        {httpnarrowwaist}
\bibfield{author}{\bibinfo{person}{Lucian Popa}, \bibinfo{person}{Ali Ghodsi}, {and} \bibinfo{person}{Ion Stoica}.} \bibinfo{year}{2010}\natexlab{}.
\newblock \showarticletitle{HTTP as the narrow waist of the future internet}. In \bibinfo{booktitle}{\emph{Proceedings of the 9th ACM SIGCOMM Workshop on Hot Topics in Networks}} (Monterey, California) \emph{(\bibinfo{series}{Hotnets-IX})}. \bibinfo{publisher}{Association for Computing Machinery}, \bibinfo{address}{New York, NY, USA}, Article \bibinfo{articleno}{6}, \bibinfo{numpages}{6}~pages.
\newblock
\showISBNx{9781450304092}
\href{https://doi.org/10.1145/1868447.1868453}{doi:\nolinkurl{10.1145/1868447.1868453}}


\bibitem[Ramanujam et~al\mbox{.}(2021)]%
        {marauder}
\bibfield{author}{\bibinfo{person}{Murali Ramanujam}, \bibinfo{person}{Harsha~V. Madhyastha}, {and} \bibinfo{person}{Ravi Netravali}.} \bibinfo{year}{2021}\natexlab{}.
\newblock \showarticletitle{Marauder: synergized caching and prefetching for low-risk mobile app acceleration}. In \bibinfo{booktitle}{\emph{Proceedings of the 19th Annual International Conference on Mobile Systems, Applications, and Services}} (Virtual Event, Wisconsin) \emph{(\bibinfo{series}{MobiSys '21})}. \bibinfo{publisher}{Association for Computing Machinery}, \bibinfo{address}{New York, NY, USA}, \bibinfo{pages}{350–362}.
\newblock
\showISBNx{9781450384438}
\href{https://doi.org/10.1145/3458864.3466866}{doi:\nolinkurl{10.1145/3458864.3466866}}


\bibitem[Rasool et~al\mbox{.}(2024)]%
        {llm_test_input}
\bibfield{author}{\bibinfo{person}{Zafaryab Rasool}, \bibinfo{person}{Scott Barnett}, \bibinfo{person}{David Willie}, \bibinfo{person}{Stefanus Kurniawan}, \bibinfo{person}{Sherwin Balugo}, \bibinfo{person}{Srikanth Thudumu}, {and} \bibinfo{person}{Mohamed Abdelrazek}.} \bibinfo{year}{2024}\natexlab{}.
\newblock \showarticletitle{LLMs for Test Input Generation for Semantic Applications}. In \bibinfo{booktitle}{\emph{Proceedings of the IEEE/ACM 3rd International Conference on AI Engineering - Software Engineering for AI}} (Lisbon, Portugal) \emph{(\bibinfo{series}{CAIN '24})}. \bibinfo{publisher}{Association for Computing Machinery}, \bibinfo{address}{New York, NY, USA}, \bibinfo{pages}{160–165}.
\newblock
\showISBNx{9798400705915}
\href{https://doi.org/10.1145/3644815.3644948}{doi:\nolinkurl{10.1145/3644815.3644948}}


\bibitem[Sathish et~al\mbox{.}(2024)]%
        {llempower}
\bibfield{author}{\bibinfo{person}{Vishwas Sathish}, \bibinfo{person}{Hannah Lin}, \bibinfo{person}{Aditya~K Kamath}, {and} \bibinfo{person}{Anish Nyayachavadi}.} \bibinfo{year}{2024}\natexlab{}.
\newblock \showarticletitle{LLeMpower: Understanding Disparities in the Control and Access of Large Language Models}.
\newblock \bibinfo{journal}{\emph{arXiv preprint arXiv:2404.09356}} (\bibinfo{year}{2024}).
\newblock


\bibitem[Shekhar et~al\mbox{.}(2024)]%
        {shekhar2024optimizingcostsllmusage}
\bibfield{author}{\bibinfo{person}{Shivanshu Shekhar}, \bibinfo{person}{Tanishq Dubey}, \bibinfo{person}{Koyel Mukherjee}, \bibinfo{person}{Apoorv Saxena}, \bibinfo{person}{Atharv Tyagi}, {and} \bibinfo{person}{Nishanth Kotla}.} \bibinfo{year}{2024}\natexlab{}.
\newblock \showarticletitle{Towards Optimizing the Costs of LLM Usage}.
\newblock \bibinfo{journal}{\emph{arXiv preprint arXiv:2402.01742}} (\bibinfo{year}{2024}).
\newblock
\showeprint{2402.01742}~[cs.CL]
\urldef\tempurl%
\url{https://arxiv.org/abs/2402.01742}
\showURL{%
\tempurl}


\bibitem[Sheng et~al\mbox{.}(2024)]%
        {fairness_osdi}
\bibfield{author}{\bibinfo{person}{Ying Sheng}, \bibinfo{person}{Shiyi Cao}, \bibinfo{person}{Dacheng Li}, \bibinfo{person}{Banghua Zhu}, \bibinfo{person}{Zhuohan Li}, \bibinfo{person}{Danyang Zhuo}, \bibinfo{person}{Joseph~E. Gonzalez}, {and} \bibinfo{person}{Ion Stoica}.} \bibinfo{year}{2024}\natexlab{}.
\newblock \showarticletitle{Fairness in Serving Large Language Models}. In \bibinfo{booktitle}{\emph{18th USENIX Symposium on Operating Systems Design and Implementation (OSDI 24)}}. \bibinfo{publisher}{USENIX Association}, \bibinfo{address}{Santa Clara, CA}, \bibinfo{pages}{965--988}.
\newblock
\showISBNx{978-1-939133-40-3}
\urldef\tempurl%
\url{https://www.usenix.org/conference/osdi24/presentation/sheng}
\showURL{%
\tempurl}


\bibitem[Shnitzer et~al\mbox{.}(2024)]%
        {shnitzer2024large}
\bibfield{author}{\bibinfo{person}{Tal Shnitzer}, \bibinfo{person}{Anthony Ou}, \bibinfo{person}{M{\'\i}rian Silva}, \bibinfo{person}{Kate Soule}, \bibinfo{person}{Yuekai Sun}, \bibinfo{person}{Justin Solomon}, \bibinfo{person}{Neil Thompson}, {and} \bibinfo{person}{Mikhail Yurochkin}.} \bibinfo{year}{2024}\natexlab{}.
\newblock \showarticletitle{Large Language Model Routing with Benchmark Datasets}. In \bibinfo{booktitle}{\emph{First Conference on Language Modeling}}.
\newblock
\urldef\tempurl%
\url{https://openreview.net/forum?id=Zb0ajZ7vAt}
\showURL{%
\tempurl}


\bibitem[Singh et~al\mbox{.}(2015)]%
        {flexiweb}
\bibfield{author}{\bibinfo{person}{Shailendra Singh}, \bibinfo{person}{Harsha~V. Madhyastha}, \bibinfo{person}{Srikanth~V. Krishnamurthy}, {and} \bibinfo{person}{Ramesh Govindan}.} \bibinfo{year}{2015}\natexlab{}.
\newblock \showarticletitle{FlexiWeb: Network-Aware Compaction for Accelerating Mobile Web Transfers}. In \bibinfo{booktitle}{\emph{Proceedings of the 21st Annual International Conference on Mobile Computing and Networking}} (Paris, France) \emph{(\bibinfo{series}{MobiCom '15})}. \bibinfo{publisher}{Association for Computing Machinery}, \bibinfo{address}{New York, NY, USA}, \bibinfo{pages}{604–616}.
\newblock
\showISBNx{9781450336192}
\href{https://doi.org/10.1145/2789168.2790128}{doi:\nolinkurl{10.1145/2789168.2790128}}


\bibitem[Tan et~al\mbox{.}(2024)]%
        {teola}
\bibfield{author}{\bibinfo{person}{Xin Tan}, \bibinfo{person}{Yimin Jiang}, \bibinfo{person}{Yitao Yang}, {and} \bibinfo{person}{Hong Xu}.} \bibinfo{year}{2024}\natexlab{}.
\newblock \showarticletitle{Teola: Towards End-to-End Optimization of LLM-based Applications}.
\newblock \bibinfo{journal}{\emph{arXiv preprint arXiv:2407.00326}} (\bibinfo{year}{2024}).
\newblock


\bibitem[Walfish et~al\mbox{.}(2004)]%
        {doa}
\bibfield{author}{\bibinfo{person}{Michael Walfish}, \bibinfo{person}{Jeremy Stribling}, \bibinfo{person}{Maxwell~N Krohn}, \bibinfo{person}{Hari Balakrishnan}, \bibinfo{person}{Robert~Tappan Morris}, {and} \bibinfo{person}{Scott Shenker}.} \bibinfo{year}{2004}\natexlab{}.
\newblock \showarticletitle{Middleboxes No Longer Considered Harmful.}. In \bibinfo{booktitle}{\emph{OSDI}}, Vol.~\bibinfo{volume}{4}. \bibinfo{pages}{15--15}.
\newblock


\bibitem[Wang et~al\mbox{.}(2016)]%
        {shandian}
\bibfield{author}{\bibinfo{person}{Xiao~Sophia Wang}, \bibinfo{person}{Arvind Krishnamurthy}, {and} \bibinfo{person}{David Wetherall}.} \bibinfo{year}{2016}\natexlab{}.
\newblock \showarticletitle{Speeding up Web Page Loads with Shandian}. In \bibinfo{booktitle}{\emph{13th USENIX Symposium on Networked Systems Design and Implementation (NSDI 16)}}. \bibinfo{publisher}{USENIX Association}, \bibinfo{address}{Santa Clara, CA}, \bibinfo{pages}{109--122}.
\newblock
\showISBNx{978-1-931971-29-4}
\urldef\tempurl%
\url{https://www.usenix.org/conference/nsdi16/technical-sessions/presentation/wang}
\showURL{%
\tempurl}


\bibitem[Yu et~al\mbox{.}(2022)]%
        {orca}
\bibfield{author}{\bibinfo{person}{Gyeong-In Yu}, \bibinfo{person}{Joo~Seong Jeong}, \bibinfo{person}{Geon-Woo Kim}, \bibinfo{person}{Soojeong Kim}, {and} \bibinfo{person}{Byung-Gon Chun}.} \bibinfo{year}{2022}\natexlab{}.
\newblock \showarticletitle{Orca: A Distributed Serving System for {Transformer-Based} Generative Models}. In \bibinfo{booktitle}{\emph{16th USENIX Symposium on Operating Systems Design and Implementation (OSDI 22)}}. \bibinfo{publisher}{USENIX Association}, \bibinfo{address}{Carlsbad, CA}, \bibinfo{pages}{521--538}.
\newblock
\showISBNx{978-1-939133-28-1}
\urldef\tempurl%
\url{https://www.usenix.org/conference/osdi22/presentation/yu}
\showURL{%
\tempurl}


\bibitem[Zhang et~al\mbox{.}(2024)]%
        {zhang2024llmcascademultiobjectiveoptimal}
\bibfield{author}{\bibinfo{person}{Kai Zhang}, \bibinfo{person}{Liqian Peng}, \bibinfo{person}{Congchao Wang}, \bibinfo{person}{Alec Go}, {and} \bibinfo{person}{Xiaozhong Liu}.} \bibinfo{year}{2024}\natexlab{}.
\newblock \bibinfo{title}{LLM Cascade with Multi-Objective Optimal Consideration}.
\newblock
\showeprint[arxiv]{2410.08014}~[cs.CL]
\urldef\tempurl%
\url{https://arxiv.org/abs/2410.08014}
\showURL{%
\tempurl}


\bibitem[Zheng et~al\mbox{.}(2023)]%
        {zheng2023judging}
\bibfield{author}{\bibinfo{person}{Lianmin Zheng}, \bibinfo{person}{Wei-Lin Chiang}, \bibinfo{person}{Ying Sheng}, \bibinfo{person}{Siyuan Zhuang}, \bibinfo{person}{Zhanghao Wu}, \bibinfo{person}{Yonghao Zhuang}, \bibinfo{person}{Zi Lin}, \bibinfo{person}{Zhuohan Li}, \bibinfo{person}{Dacheng Li}, \bibinfo{person}{Eric~P. Xing}, \bibinfo{person}{Hao Zhang}, \bibinfo{person}{Joseph~E. Gonzalez}, {and} \bibinfo{person}{Ion Stoica}.} \bibinfo{year}{2023}\natexlab{}.
\newblock \showarticletitle{Judging LLM-as-a-Judge with MT-Bench and Chatbot Arena}.
\newblock \bibinfo{journal}{\emph{arXiv preprint arXiv:2306.05685}} (\bibinfo{year}{2023}).
\newblock
\showeprint{2306.05685}~[cs.CL]


\bibitem[Zhu et~al\mbox{.}(2024)]%
        {zhu2024efficientpromptcachingembedding}
\bibfield{author}{\bibinfo{person}{Hanlin Zhu}, \bibinfo{person}{Banghua Zhu}, {and} \bibinfo{person}{Jiantao Jiao}.} \bibinfo{year}{2024}\natexlab{}.
\newblock \showarticletitle{Efficient Prompt Caching via Embedding Similarity}.
\newblock \bibinfo{journal}{\emph{arXiv preprint arXiv:2402.01173}} (\bibinfo{year}{2024}).
\newblock
\showeprint{2402.01173}~[cs.CL]
\urldef\tempurl%
\url{https://arxiv.org/abs/2402.01173}
\showURL{%
\tempurl}


\end{thebibliography}

%%%%%%%%%%%%%%%%%%%%%%%%%%%%%%%%%%%%%%%%%%%%%%%%%%%%%%%%%%%%%%%%%%%%%%%%%%%%%%%%
\end{document}